\documentclass[aps,twocolumn,prd,nofootinbib,superscriptaddress,preprintnumbers]{revtex4-2}

\usepackage[utf8]{inputenc}

\usepackage{mathtools}
\usepackage{amsfonts}
\usepackage{amssymb}
\usepackage{mathrsfs}
\usepackage{bbm}
\usepackage{slashed}
\usepackage{tensor}

\usepackage{graphicx}
\usepackage{array}

\usepackage{placeins}
\usepackage{makecell}
\usepackage[caption=false]{subfig}
\usepackage{float}

\usepackage{xspace}
\usepackage{xfrac}
\usepackage{hyperref}
\usepackage[nameinlink]{cleveref}
\usepackage{appendix}
\usepackage{units}

\usepackage{xifthen}
\usepackage{booktabs}
\usepackage{multirow}
\usepackage{courier}
\usepackage[dvipsnames]{xcolor}
\hypersetup{
	colorlinks,
	linkcolor={blue!75!black},
	citecolor={blue!75!black},
	urlcolor={blue!75!black}
}

\newcommand{\rmd}{\mathrm{d}}
\newcommand{\rmi}{\mathrm{i}}
\newcommand{\rmSU}{\mathrm{SU}}

\graphicspath{{./figs/}}

\newcommand{\gettitle}{Normalizing flows for lattice gauge theory in arbitrary space-time dimension}

\newcommand{\getMITAffiliation}{\affiliation{Center for Theoretical Physics, Massachusetts Institute of Technology, Cambridge, MA 02139, USA}}
\newcommand{\getIAIFIAffiliation}{\affiliation{The NSF AI Institute for Artificial Intelligence and Fundamental Interactions}}
\newcommand{\getNYUAffiliation}{\affiliation{Center for Cosmology and Particle Physics, New York University, New York, NY 10003, USA}}
\newcommand{\getDMAffiliation}{\affiliation{DeepMind, London, UK}}
\newcommand{\getWisconsinAffiliation}{\affiliation{Physics Department, University of Wisconsin-Madison, Madison, WI 53706, USA}}
\newcommand{\getBernAffiliation}{\affiliation{Albert Einstein Center, Institute for Theoretical Physics, University of Bern, 3012 Bern, Switzerland}}

\hypersetup{
    pdftitle={\gettitle},
    pdfauthor={Abbott,Albergo,Botev,Boyda,Cranmer,Hackett,Kanwar,Matthews,Racanière,Razavi,Rezende,Romero-López,Shanahan,Urban},
    pdfkeywords={lattice field theory}{lattice gauge theory}{normalizing flows}{machine learning},
    bookmarksopen=true,
    bookmarksopenlevel=2,
    bookmarksnumbered=true
}

\begin{document}

\title{\gettitle}

\author{Ryan~Abbott}
\getMITAffiliation
\getIAIFIAffiliation
\author{Michael~S.~Albergo}
\getNYUAffiliation
\author{Aleksandar Botev}
\getDMAffiliation
\author{Denis~Boyda}
\getMITAffiliation
\getIAIFIAffiliation
\author{Kyle~Cranmer}
\getWisconsinAffiliation
\author{Daniel~C.~Hackett}
\getMITAffiliation
\getIAIFIAffiliation
\author{Gurtej~Kanwar}
\getBernAffiliation
\getMITAffiliation
\getIAIFIAffiliation
\author{Alexander~G.D.G.~Matthews}
\getDMAffiliation
\author{S\'{e}bastien~Racani\`{e}re}
\getDMAffiliation
\author{Ali Razavi}
\getDMAffiliation
\author{Danilo~J.~Rezende}
\getDMAffiliation
\author{Fernando~Romero-L\'opez}
\getMITAffiliation
\getIAIFIAffiliation
\author{Phiala~E.~Shanahan}
\getMITAffiliation
\getIAIFIAffiliation
\author{Julian~M.~Urban}
\getMITAffiliation
\getIAIFIAffiliation

\preprint{MIT-CTP/5554}

\begin{abstract}
    Applications of normalizing flows to the sampling of field configurations in lattice gauge theory have so far been explored almost exclusively in two space-time dimensions. We report new algorithmic developments of gauge-equivariant flow architectures facilitating the generalization to higher-dimensional lattice geometries. Specifically, we discuss masked autoregressive transformations with tractable and unbiased Jacobian determinants, a key ingredient for scalable and asymptotically exact flow-based sampling algorithms. For concreteness, results from a proof-of-principle application to SU(3) lattice gauge theory in four space-time dimensions are reported.
\end{abstract}

\maketitle

\section{Introduction}\label{sec:intro}

The trivializing map hybrid Monte Carlo (HMC) algorithm~\cite{Luscher:2009eq} has inspired many recent efforts~\cite{Albergo:2019eim, Kanwar:2020xzo, Nicoli:2020njz, Boyda:2020hsi, Albergo:2021vyo, Albergo:2021bna, DelDebbio:2021qwf, Hackett:2021idh, Nicoli:2021inv, Foreman:2021ljl, Finkenrath:2022ogg, Albergo:2022qfi, Pawlowski:2022rdn, Gerdes:2022eve, Singha:2022lpi, Abbott:2022zhs, Abbott:2022hkm, Abbott:2022zsh, Bacchio:2022vje, Komijani:2023fzy, Albandea:2023wgd, Nicoli:2023qsl, R:2023dcr} to construct machine-learned generative models for asymptotically exact sampling in lattice field theory, primarily based on normalizing flows (NF)~\cite{tabak2010, tabak2013, dinh2015nice, rezende2016variational, dinh2017density}. Apart from other promising areas of application such as the computation of thermodynamic observables, this line of research aims to mitigate prohibitive ergodicity problems like critical slowing-down and topological freezing in lattice quantum chromodynamics (LQCD) calculations. Addressing these issues would greatly benefit various applications spanning from high-precision studies of proton structure to first-principles calculations of nuclei~\cite{Lehner:2019wvv, Kronfeld:2019nfb, Cirigliano:2019jig, Detmold:2019ghl, Bazavov:2019lgz, Joo:2019byq}. So far, proof-of-principle studies of flow-based sampling for lattice gauge theory (LGT) have been carried out primarily in two space-time dimensions. Efficient mixing of topological sectors has been demonstrated in settings including $\mathrm{U}(1)$ pure gauge theory~\cite{Kanwar:2020xzo} and the Schwinger model~\cite{Finkenrath:2022ogg, Albergo:2022qfi}. For the application to LQCD~\cite{Abbott:2022hkm}, the development of scalable flow models for four-dimensional LGT with gauge group $\rmSU(3)$ is a crucial step.

In this work, we consider NF architectures applicable to non-Abelian LGT with in-principle arbitrary space-time lattice geometry. After discussing general structural properties and requirements of scalable flow models, gauge-equivariant flows based on coupling layers~\cite{dinh2015nice} are introduced. Specifically, we consider conditional element-wise transformations of subsets of variables with decompositions generated by masking patterns, enabling both the probabilistic modeling of optimal context-dependent maps via machine learning as well as the efficient and unbiased computation of Jacobian determinants. These features are likely to be essential ingredients for the construction of asymptotically exact flow-based sampling algorithms at scale. Here, we focus on \emph{spectral} flows transforming eigenvalue spectra of untraced Wilson loops, as well as gradient-based \emph{residual} flows directly transforming gauge links based on derivatives of potential functions. Both types of flows may also be freely combined into arbitrary stacks of layers. A variety of possible design choices emerge at different levels of abstraction, providing a basis for the systematic exploration of model architectures. We report general observations about the advantages and disadvantages of different components, such as training strategies and neural network parameterizations, as well as a number of low-level implementation details. These developments are demonstrated in a numerical example for $\rmSU(3)$ LGT in (3+1)d.

This paper is organized as follows. In \Cref{sec:background}, aspects of LGT and NFs are reviewed. In \Cref{sec:trfs}, the general structure of scalable NFs for LGT is discussed, followed by concrete constructions in \Cref{sec:coupling}. Neural network parameterizations and training strategies are explored in \Cref{sec:context}. Numerical results are reported in \Cref{sec:demo}. \Cref{sec:summary} provides a summary and outlook.

\section{Background}\label{sec:background}

\subsection{Lattice gauge theory}

The goal of Markov chain Monte Carlo (MCMC) sampling in lattice field theory is to obtain field configurations $U$ defined on a discrete $d$-dimensional Euclidean space-time lattice and distributed according to a target density $p(U) = \frac{1}{Z} e^{-S(U)}$. Here, $S$ is the lattice action of the theory under consideration, and the normalization or partition function $Z = \int \rmd U \, e^{-S(U)}$ is usually unknown. Ensembles of field configurations may then be used to estimate path integral expectation values of quantum operators in thermodynamic equilibrium. Importantly, these integrals are often of extremely high dimensionality, not to be confused with the space-time dimension: with $L$ lattice sites in each direction, the number of integration variables is generally $\mathcal{O}(L^d)$, with the precise value depending on the internal structure and geometric properties of the considered quantum field. In some state-of-the-art calculations, each configuration is represented by $\mathcal{O}(10^{11})$ double-precision floating point numbers~\cite{Fritzsch:2021klm}.

For LGT, the degrees of freedom of a field configuration are group elements $U_\mu(x) \in G$ defined on the links between lattice sites, where $x,\mu$ denote the position and direction of the corresponding link and one usually assumes periodic boundary conditions in all directions. Restricting to gauge group $\rmSU(N)$, the simplest lattice action permitting a consistent continuum limit is the Wilson action~\cite{Wilson:1974sk},
\begin{equation}\label{eq:action}
    S(U) = -\frac{\beta}{N} \sum_{x} \sum_{\substack{\mu,\nu \\ \mu < \nu}} \mathrm{Re}\, \mathrm{Tr}\, {P_{\mu\nu}(x)}
\,.
\end{equation}
Here, $\beta$ is proportional to the inverse of the squared gauge coupling, and the plaquette $P_{\mu\nu}(x)$ is the smallest possible Wilson loop---a product of gauge links around a $1 \times 1$ square, defined as
\begin{equation}
    P_{\mu\nu}(x) = U_\mu(x) U_\nu(x+\hat{\mu}) U^\dagger_{\mu}(x+\hat{\nu}) U^\dagger_{\nu}(x)
\,.
\end{equation}
The action is invariant under local gauge transformations of the form
\begin{equation}\label{eq:gauge}
    g: U_\mu(x) \longmapsto \Omega(x) U_\mu(x) \Omega^\dagger(x+\hat{\mu})
\,,
\end{equation}
where $\Omega(x) \in G$ can be chosen arbitrarily and independently at every lattice site $x$.

Since its inception as a numerical approach towards non-perturbative QCD, the most common sampling algorithms for LGT have been HMC~\cite{DUANE1987216} and \mbox{(pseudo-)heatbath} with overrelaxation steps~\cite{Creutz:1980zw, Cabibbo:1982zn, Kennedy:1985nu, Brown:1987rra, Adler:1987ce}. However, ergodicity issues such as topological freezing and critical slowing-down are encountered at small lattice spacings~\cite{Schaefer:2009xx}. These problems generally lead to large autocorrelation times in Markov chains generated by local or diffusive sampling algorithms and currently limit the precision and reach of lattice studies of many interesting problems, due to the explosion of the computational cost required to obtain statistically independent field configurations as the continuum limit is approached.

\subsection{Normalizing flows for lattice field theory}

In recent years, asymptotically exact samplers for lattice field theories have been constructed using machine-learned models that allow direct generation of independent field configurations $U$ with a known and tractable probability density $q(U)$, optimized to approximate a desired target density $p(U)$. The simplest and most direct sampling approach in this context is the independence Metropolis algorithm~\cite{Metropolis:1953am, tierney1994markov}: candidate samples are first generated by embarrassingly parallel evaluation of the model, and then used to construct a Markov chain in a computationally inexpensive post-processing step. Asymptotic exactness of the stationary distribution is achieved by proposing transitions $U \rightarrow U'$ and accepting them with probability
\begin{equation}
\begin{aligned}
    A &= \min\left(1,\frac{p(U')}{p(U)}\frac{q(U)}{q(U')}\right) \\
      &= \min\left(1,\frac{e^{-S(U') - \log q(U')}}{e^{-S(U) - \log q(U)}}\right)
\,.
\end{aligned}
\end{equation}
This algorithm has the advantage that, by construction, every accepted field configuration fully decorrelates the chain due to being statistically independent a priori. Hence, autocorrelations are generated entirely by rejections, and perfect mixing is achieved in the ideal limit of perfect models, $q(U) = p(U)$. However, many different sampling schemes with potentially favorable statistical properties can be devised, including hybrid approaches that combine generative models with more traditional algorithms~\cite{Foreman:2021ljl, Albandea:2023wgd}.

NFs represent one particular family of generative machine learning models with computationally tractable probability densities. So far, flow-based algorithms for the sampling of lattice field configurations have been developed in the context of (mostly two-dimensional) \mbox{$\phi^4$-theory}, $\mathrm{U}(1)$ and $\rmSU(N)$ gauge theories, as well as theories containing fermions~\cite{Albergo:2019eim, Kanwar:2020xzo, Nicoli:2020njz, Boyda:2020hsi, Albergo:2021vyo, Albergo:2021bna, DelDebbio:2021qwf, Hackett:2021idh, Nicoli:2021inv, Foreman:2021ljl, Finkenrath:2022ogg, Albergo:2022qfi, Pawlowski:2022rdn, Gerdes:2022eve, Singha:2022lpi, Abbott:2022zhs, Abbott:2022hkm, Abbott:2022zsh, Bacchio:2022vje, Komijani:2023fzy, Albandea:2023wgd, Nicoli:2023qsl, R:2023dcr}. First results for flow-based sampling applied to LQCD in four space-time dimensions have been reported recently~\cite{Abbott:2022hkm}, already utilizing some of the novel algorithmic developments introduced in the present work. Aspects of scaling and scalability have also been discussed in detail in Ref.~\cite{Abbott:2022zsh}. For an introduction to NFs for lattice field theory including a working implementation, see Ref.~\cite{Albergo:2021vyo}.

A flow-based model consists of a diffeomorphism $f$ and a prior density $r(\cdot)$, which can together be used to produce samples by first drawing $V$ according to $r(V)$ and then evaluating $U = f(V)$. The Jacobian $J = \frac{\partial f}{\partial V}$, combined with the prior density, allows the output probability density $q$ to be evaluated as
\begin{equation}
    q(U) = r(V) \left|\det J \right|^{-1}
\,.
\end{equation}
Similar to many modern neural network architectures, a flow is commonly decomposed into a sequence of layers or steps, with the crucial difference that these elementary transformations must all be invertible. The components are parameterized by free model parameters (`weights'), which can be numerically optimized to minimize the difference between the model density $q$ and target density $p$. An observable-independent indicator of the model quality is given by the effective sample size per configuration (ESS), which quantifies the effective fraction of independent samples obtained after accounting for autocorrelations (or reweighting variance) when using $q$ to sample $p$. It is defined as the inverse of the average squared reweighting factor $w = p/q$ and can be statistically estimated from a batch of $K$ model samples $U_i \sim q(U_i)$,
\begin{equation}
    \text{ESS} = \frac{1}{\mathbb{E}_q[w^2]} \approx \frac{\left(\nicefrac{1}{K} \sum_i \tilde{w}(U_i) \right)^2}{\nicefrac{1}{K} \sum_i \tilde{w}(U_i)^2} \in \left[\frac{1}{K}, 1\right]
\,,
\end{equation}
where $\tilde{w}(U_i) = \exp[-S(U_i)] / q(U_i)$. Larger values correspond to better models, and $\text{ESS} = 1$ in the limit of identical model and target densities. The ESS is a useful metric in this context because it can be computed from model samples alone, without the need to build a Markov chain to determine acceptance rates.

The free parameters of a NF, or more generally any parameterized diffeomorphism defined on the manifold of field variables, may be optimized by minimizing a suitable loss function. In particular, assuming that stochastic estimates of a given loss function are differentiable with respect to the considered parameters, optimization based on gradient descent can be applied, as is standard for the training of large neural networks~\cite{LeCuBottOrrMull9812}. In the context of optimizing NFs to approximate an unnormalized target density, a commonly used loss function is the (conventionally) \emph{reverse} Kullback-Leibler divergence (KLD)~\cite{Kullback:1951}, a measure of the relative entropy between $q$ and $p$. It is defined as
\begin{equation}\label{eq:kld}
\begin{aligned}
    D_{\mathrm{KL,rev}} &= \int \rmd U \, q(U) \log \frac{q(U)}{p(U)} \\
    &= \mathbb{E}_{U \sim q(U)} \Big[\log q(U) + S(U)\Big] + \log Z
\,,
\end{aligned}
\end{equation}
where the constant $\log Z$ arising from the unknown normalization of the target $p$ is irrelevant for the purpose of training as it does not affect the gradients, and can be formally subtracted. However, the absolute meaning of the numerical value of the loss as a measure of the difference between $q$ and $p$ is then lost---rather, it now provides a bound on $\log Z$. Importantly, estimates are defined with respect to the model density $q$. No samples from the target $p$ are required for the optimization; hence, the approach is fully variational.

While one often considers directly sampleable prior distributions with fully tractable and normalized densities $r$ (such as a standard normal for scalar fields or the Haar measure for LGT), this is not a necessary condition for consistent training and sampling of a flow-based model. In fact, the minimal requirements are merely that the prior can be sampled and that unnormalized prior and target densities can be computed per sample, implying greater flexibility in the choice of $r$. For instance, it is possible to define flow transformations between distributions that only differ in the values of parameters of the lattice action, thereby more closely following the interpretation of trivializing and/or optimal transport maps as renormalization group flows~\cite{Luscher:2009eq, Cotler:2022fze}. This is often computationally more efficient since the transformations can naturally be parameterized more simply if the prior and target distributions are more similar. For LGT in particular, it may be advantageous to use a prior with the same type of action as the target, but smaller values of $\beta$ or different boundary conditions such that topological sectors mix sufficiently fast using traditional sampling algorithms, or to transform pure LGT with infinite quark masses to LQCD with dynamical quarks (quenched to unquenched). While such strategies necessarily sacrifice some of the advantages of independent sampling with fully known model densities, they may still provide significant gains in efficiency compared to sampling the target directly. Moreover, if the flow is employed merely as a change-of-variables embedded within a diffusive algorithm like HMC~\cite{Albandea:2023wgd} (as in the original trivializing map proposal~\cite{Luscher:2009eq}), the prior density is defined as the reverse-flowed target and does not need to be specified explicitly.

The above considerations also imply that NFs have promising applications not only in accelerated sampling, but more generally for problems where optimizable field transformations with tractable Jacobians can provide advantages. For instance, they may conceivably extend the reach of ensemble reweighting to different action parameters, aid in variance reduction for observables plagued by severe signal-to-noise issues~\cite{Detmold:2021ulb, Yunus:2022pto, Yunus:2022wdr}, as well as provide an approach towards solving thermodynamic integration problems~\cite{Nicoli:2020njz, Nicoli:2021inv, Pawlowski:2022rdn, Nicoli:2023qsl}.

\section{Scalable flows for LGT}\label{sec:trfs}

Due to the particular symmetries and geometry of LGT, constructing suitable flow models for this application is more involved than for, e.g., image data or scalar fields. For one, the relevant degrees of freedom are not just real numbers living on the sites of a lattice or the pixels of an image, but are group-valued links connecting neighboring sites. For the common case of a compact Lie group, implementing appropriate transformations requires parameterized functions defined on compact manifolds; see also Ref.~\cite{Rezende:2020hrd}. Furthermore, the high dimensionality of gauge field configurations in state-of-the-art calculations implies very stringent scalability constraints and corresponding engineering efforts.

In this section, we first consider a suitable set of requirements for NF architectures that facilitate their scalable application to LGT. We then discuss some basic design principles for the parameterization of transformations that are both expressive and computationally tractable, based on organizing flows into coupling layers.

\subsection{Structure and requirements}

\begin{figure}
    \centering
    \includegraphics[width=\linewidth]{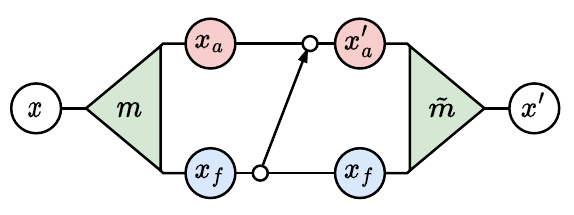}
    \caption{Structure of a masked autoregressive flow transformation with triangular Jacobian, i.e., a coupling layer. First, a set of variables $x$ is decomposed into active and frozen subsets $x_a,x_f$ using a mask $m$. The active variables are then transformed element-wise to $x_a'$, conditioned on $x_f$. The active and frozen subsets are then re-combined into $x'$ by reversing the initial decomposition, written symbolically as $\tilde m$. In order to construct expressive flow transformations, many such layers with alternating masking patterns are combined sequentially.}
    \label{fig:coupling-layer}
\end{figure}

For the prospective application of NFs to at-scale LQCD calculations, the cost of evaluating the model itself must obviously remain tractable. While it is still unclear whether high-quality models can eventually be obtained for the volumes and parameters of interest, a necessary condition is the naive scalability of all components of the flow transformation; see also Ref.~\cite{Abbott:2022zsh}. In other words, while the performance \emph{scaling} of different architectures needs to be determined empirically, constructing them to be \emph{scalable} a priori without an intrinsically unfavorable growth of the computational cost is a useful design principle. This concerns in particular the computation of the Jacobian determinant: the cost naively grows cubically in the number of variables unless restrictions are imposed on the transformation that simplify the structure of the matrix. In the present work, we only consider such suitably restricted flow layers. While tractable and asymptotically exact sampling may in principle also be achieved via unbiased stochastic estimates of the determinant, the treatment of fermion determinants in LQCD provides a cautionary tale in this regard~\cite{Jansen:2003nt}, suggesting that the additional variance introduced by direct estimation may be prohibitive for the system sizes of interest.

In addition to the naive scalability requirement, it is often advantageous to incorporate the symmetries of the target distribution into the design of the flow architecture to some extent, in order to avoid modeling physically redundant information. For the particular case of gauge symmetries, flows can be designed to exhibit gauge \emph{equivariance}, meaning that $f$ commutes with local gauge transformations $g$ as defined in \Cref{eq:gauge}, 
\begin{equation}
    f(g \circ U) = g \circ f(U)
\,.
\end{equation}
If the prior density $r$ is gauge-invariant, gauge equivariance of the flow implies gauge invariance of the model. Of course, explicit breaking of target symmetries in the model density does not generally violate statistical exactness and is done routinely in the case of discrete space-time lattice symmetries like translation and rotation, due to necessary trade-offs in the model structure. However, the gauge symmetry considered here is continuous and high-dimensional, and implementing equivariance was found to be crucial for non-trivial sampling performance in previous studies~\cite{Kanwar:2020xzo, Boyda:2020hsi}. Gauge-equivariant neural networks~\cite{Favoni:2020reg, Nagai:2021bhh, Namekawa:2022liz, Favoni:2022mcg, Aronsson:2023rli} are also being investigated for other purposes, such as for the design of optimal preconditioners in fermion matrix inversions~\cite{Lehner:2023bba, Lehner:2023prf}. In general, matching the symmetries of a problem as closely as possible often proves to be important in the construction of expressive neural network architectures for various applications~\cite{satorras2022en, Reh:2023zzh}. In the context of flows, this approach may also avoid or expose issues of mode collapse that are otherwise difficult to detect, and ensures the invariance of reweighted expectation values---involving reweighting factors of the form $p(U_i)/q(U_i)$---under symmetry transformations that leave $p$ unchanged.

\subsection{Tractable Jacobians via masking}\label{sec:jacobian}

Tractability of the Jacobian determinant can be achieved by organizing the flow into coupling layers with element-wise transformations of suitable subsets of all variables in each layer, conditioned on the remaining variables such that the Jacobian becomes triangular. The computational complexity of the determinant then grows only linearly in the volume. The partitioning can be implemented via masking patterns that define which variables are changed (`active') conditioned on the unchanged (`frozen') variables. The basic structure of a coupling layer is illustrated in \Cref{fig:coupling-layer}.

For concreteness, consider a set of four variables $a,b,c,d$ with $a,b$ active and $c,d$ frozen, and conditional element-wise transformations of the form
\begin{equation}
    a \longmapsto a'(a|c,d) \quad \text{and} \quad b \longmapsto b'(b|c,d)
\,.
\end{equation}
The Jacobian matrix of the map $a,b,c,d \longmapsto a',b',c,d$ is then triangular,
\begin{equation}\label{eq:coupling-jacobian}
    J = \begin{pmatrix}
    \nicefrac{\partial a'}{\partial a} & 0 & \nicefrac{\partial a'}{\partial c} & \nicefrac{\partial a'}{\partial d} \\
    0 & \nicefrac{\partial b'}{\partial b} & \nicefrac{\partial b'}{\partial c} & \nicefrac{\partial b'}{\partial d} \\
    0 & 0 & 1 & 0 \\
    0 & 0 & 0 & 1
    \end{pmatrix}
\,,
\end{equation}
and its determinant is simply the product of all diagonal matrix elements, $\det{J} = \nicefrac{\partial a'}{\partial a} \cdot \nicefrac{\partial b'}{\partial b}$. This feature is also retained in autoregressive maps with further conditioning on the outputs of previous transformations within the same layer---e.g., $a \longmapsto a'(a|c,d,b')$---thereby `filling the triangle'. While this still avoids the aforementioned cubic cost scaling, full autoregression remains generically expensive due to its sequential nature, and breaks translation equivariance completely unless applied to only the internal degrees of freedom of a localized field variable. In the present work, we only consider masked autoregressive layers taking the form of coupling layers.

Conditional transformations with tractable Jacobians can in principle also be implemented by introducing a set of auxiliary fields to be marginalized a posteriori. However, this comes at the cost of a generic reduction in sampling performance simply due to the growth of dimensionality, which must be compensated through increased model quality. While there are certain advantages to auxiliary variables~\cite{wehenkel2020say, lee2021universal}, in this work we only consider models where tractability is achieved by masking.

The design of appropriate masking patterns ideally takes into account the interaction structure of the targeted field theory, since the masks control the propagation of information between different parts of the lattice. One example of a physically motivated mask $m(x)$ for fields defined on the lattice sites $x$ with nearest-neighbor interactions is a simple checkerboard pattern, since conditional information for the transformation of an active variable can then be gathered from its immediate vicinity; see Ref.~\cite{Albergo:2021vyo} for a concrete implementation. For LGT, each flow layer features a mask $m_\mu(x)$ covering all links, and we use the convention
\begin{equation}
    m_\mu(x) = \begin{cases}
    1 \quad \text{if} \ U_\mu(x) \ \text{is \emph{active}, and} \\
    0 \quad \text{if} \ U_\mu(x) \ \text{is \emph{frozen}.}
    \end{cases}
\end{equation}
The masks are alternated within the flow such that there are no `blind spots' and every field variable is changed at least once. In general, masking explicitly breaks the discrete translation symmetry to a smaller subgroup, even if the elementary transformations are otherwise fully equivariant. Concrete implementations of masking patterns are discussed in detail in \Cref{sec:spectral-mask,sec:residual-mask}.

\subsection{Context awareness}

The conditional element-wise transformations applied after the decomposition into active and frozen subsets are functions defined on the space of field variables; i.e., compact manifolds in the particular case of LGT with a compact gauge group. By choosing an appropriate parameterization, these functions can then be numerically optimized during training. While it is in principle also possible to just learn fixed weights for the transformations of the active variables---i.e., dropping the conditional dependence of $a',b'$ on $c,d$ in the example of the previous section, thereby rendering the Jacobian matrix in \Cref{eq:coupling-jacobian} diagonal---such a map is by construction agnostic to correlations, which is too restrictive for any but the simplest of modeling tasks. Clearly, achieving useful expressivity in flow transformations for non-trivial lattice field theory distributions with finite correlation length requires some level of awareness of the field values in localized regions around the active variables, in particular for large system sizes. This can be implemented by dynamically computing the parameters of each transformation in a context-dependent fashion.

A promising approach for the implementation of context awareness is via neural network context functions $nn(\cdot)$ that utilize frozen information in order to approximate the optimal choice of parameters of the active variable transformation in a given layer, written schematically as $a'(a|nn(c,d))$. While the theoretically most expressive flows may require access to all variables instead of merely a subset, they are fundamentally in opposition to a tractable computation of the Jacobian determinant unless other restrictions are imposed. This is precisely the point where machine learning techniques may provide a significant advantage, namely by improving the utilization of incomplete information. Some concrete instances of neural network parameterizations are discussed in \Cref{sec:context-networks}.

\section{Constructing gauge-equivariant coupling flows}\label{sec:coupling}

We now detail explicit constructions of scalable coupling flows for LGT based on the requirements discussed above. We identify two distinct families of gauge-equivariant transformations, operating on untraced Wilson loops as well as directly on the gauge links, respectively. Here, we focus on spectral and gradient-based residual flows as concrete instances. Their lower-level components, however, can often be constructed in a number of different ways, resulting in a rich landscape of possible combinations. We emphasize the points at which alternative design choices exist and provide general observations about their utility from our numerical experiments. Some implementation details and training strategies applicable to both types of flows are discussed separately in \Cref{sec:context}; concrete results for both model families are reported in \Cref{sec:demo}.

\subsection{Loop-level flow}\label{sec:loop-flow}

One possibility to parameterize a gauge-equivariant flow is via coupling layers transforming untraced Wilson loops, which has been previously studied in two space-time dimensions~\cite{Kanwar:2020xzo, Boyda:2020hsi}. In the present work, we consider the generalization of this architecture to arbitrary dimensions and discuss novel algorithmic developments, some of which have already been utilized in Ref.~\cite{Abbott:2022hkm}. We summarize first the main concepts and then describe the concrete implementation based on transforming the eigenvalue spectra of untraced loops.

Under a local gauge transformation as defined in \Cref{eq:gauge}, an untraced Wilson loop starting and ending at position $x$ transforms as
\begin{equation}
    W(x) \longmapsto \Omega(x) W(x) \Omega(x)^\dag
\,,
\end{equation}
i.e., conjugation by arbitrary elements of the gauge group. Hence, gauge equivariance of a transformation $f: W(x) \longmapsto W'(x)$ takes the form of conjugation equivariance, requiring
\begin{equation}
    f(\Omega(x) W(x) \Omega(x)^\dag) = \Omega(x) f(W(x)) \Omega(x)^\dag
\,.
\end{equation}
A gauge-equivariant map for the links is then obtained by pushing the difference between the original and transformed loop onto the first of its constituent links at position $x$,
\begin{equation}
    U_\mu(x) \longmapsto W'(x) W(x)^\dag U_\mu(x)
\,.
\end{equation}

Essentially arbitrary active loop shapes may be used to define the transformation. Wilson lines connecting field variables that transform like spinors can in principle also be employed, but are not considered in the present work. Context awareness is implemented by supplying gauge-invariant quantities (i.e., traced loops) as conditional information. For instance, one may choose to transform untraced plaquettes conditioned on any number of traced loops. To satisfy tractability of the Jacobian determinant, in addition to the partitioning of the field degrees of freedom into active and frozen subsets, the same needs to be done for the chosen loop structures. Changing a given active link also changes the values of any loops that contain it, even if they are not active. The unavoidable existence of such `passive' loops---which are not actively transformed but simultaneously cannot be utilized as frozen information---complicates the construction of appropriate masking patterns; see \Cref{sec:spectral-mask}.

\subsubsection{Spectral flow}\label{sec:spectral-flow}

The eigenvalue spectrum of an untraced Wilson loop encodes the entirety of its physical, gauge-invariant information content. Hence, one may argue that all loop-level coupling flows as described above are necessarily spectral flows, since any conjugation-equivariant map of an untraced loop must at its core operate on the spectrum. Retaining the semantic distinction simply leaves open the possibility that such flows could also be constructed without directly referring to the eigenvalues. However, in this work, we only consider spectral flows based on explicit diagonalization, as introduced in Ref.~\cite{Boyda:2020hsi}. We briefly review the basic structure and then elaborate on an improved parameterization of the transformation.

\begin{figure}
    \centering
    \includegraphics[width=\linewidth]{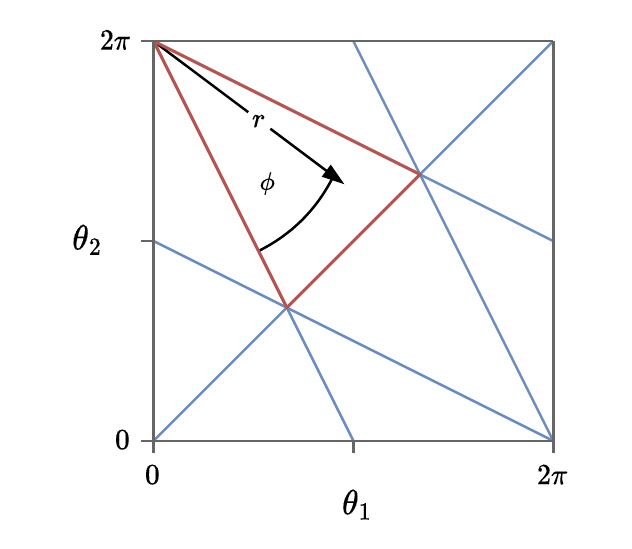}
    \caption{Depiction of the choice of canonical cell and polar coordinates used to define permutation-equivariant transformations of eigenvalue spectra of $\rmSU(3)$ matrices, resulting in conjugation-equivariant transformations of the matrices themselves and thereby implementing gauge equivariance if the matrices are taken to be untraced Wilson loops. Note that this convention for the canonical cell differs from the one used in Ref.~\cite{Boyda:2020hsi}, in order to set the origin of the polar coordinate system introduced here to the point $(0,0)$.}
    \label{fig:canonical-cell}
\end{figure}

The following procedure can be generalized to $\rmSU(N)$ with arbitrary $N$; however, for concreteness we focus solely on $\rmSU(3)$ which is most relevant for LQCD applications. In that case, the eigenvalues can be written in terms of two angles $\theta_1,\theta_2 \in [0,2\pi]$ as
\begin{equation}
    \lambda_1 = e^{i\theta_1},\ \lambda_2 = e^{i\theta_2},\ \lambda_3 = e^{-i\theta_1-i\theta_2}
\,.
\end{equation}
The $(\theta_1,\theta_2)$-plane can be divided into six cells related by the permutation group $S_3$ acting on $(\lambda_1,\lambda_2,\lambda_3)$. The requirement of gauge equivariance can be reduced to permutation equivariance of a map of the eigenvalues, with transformations defined by a three-step procedure:
\begin{enumerate}
    \item Diagonalization $U = V \Lambda V^\dag$, with $\Lambda$ the diagonal matrix containing the spectrum.
    \item Permutation-equivariant spectral flow, $\Lambda \longmapsto \Lambda'$.
    \item Re-application of the eigenvector matrices to obtain the transformed link, $U' = V \Lambda' V^\dag$.
\end{enumerate}
Equivariance under discrete groups, such as permutations, can often be realized by explicitly summing or averaging over the whole group. This symmetrization approach may be used, e.g., to achieve equivariance under $\mathbb{Z}_2$ for $\phi^4$-theory in the broken phase, or to implement translation equivariance with anti-periodic boundary conditions~\cite{Albergo:2021bna}. The present case is slightly more complicated since one needs to compute the Jacobian matrix for the map $(\theta_1,\theta_2) \longmapsto (\theta_1',\theta_2')$ but the permutation group acts on $(\lambda_1,\lambda_2,\lambda_3)$; nevertheless, no major conceptual difficulties arise. However, as already pointed out in Ref.~\cite{Boyda:2020hsi}, this construction appears to be overly restrictive for the expressivity of the transformation. Instead, we opt for a canonical ordering of the eigenvalues on which the transformation is applied, followed by the reversal of said ordering. The choice of ordering is equivalent to selecting one of the aforementioned six cells that are related by the permutation group, which will be referred to as the canonical cell; our choice is illustrated by the red triangle in \Cref{fig:canonical-cell}. This procedure is permutation-equivariant by construction, but has the advantage that the context functions do not need to disentangle the different possible permutations, potentially allowing more efficient modeling of correlations.

\begin{figure*}
    \centering
    \subfloat[]{%
        \includegraphics[width=0.332\textwidth]{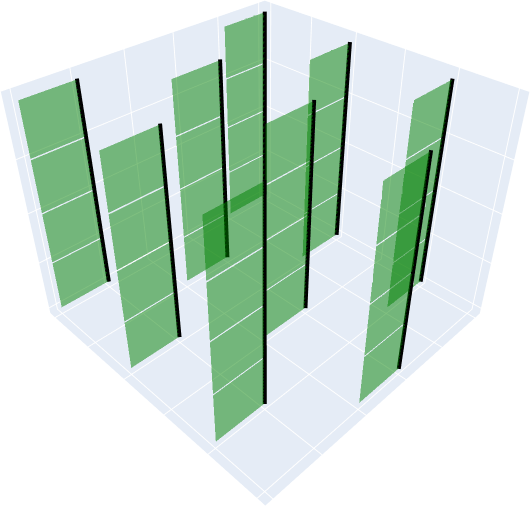}\label{fig:mask1}}%
    \hfill
    \subfloat[]{%
        \includegraphics[width=0.332\textwidth]{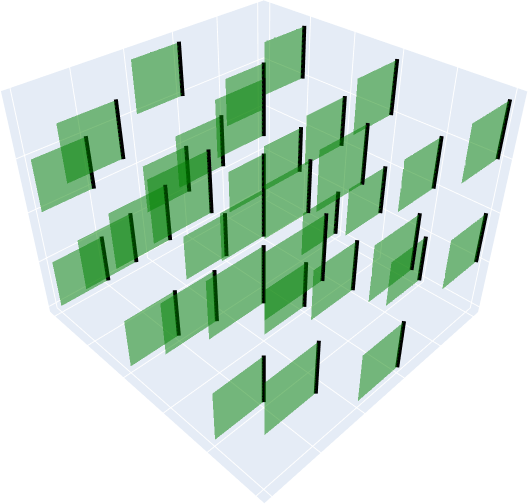}\label{fig:mask2}}%
    \hfill
    \subfloat[]{%
        \includegraphics[width=0.3\textwidth]{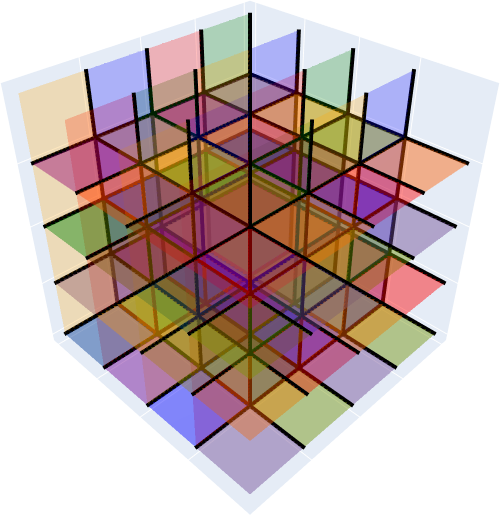}\label{fig:mask-altern}}%
    \caption{Examples of three-dimensional masking patterns generated with the algorithm described in \Cref{sec:spectral-mask} and \Cref{app:mask}. Active links are depicted as black lines and active plaquettes as colored squares. In example (b), every active link is completely surrounded by frozen links. Masks with this property generally result in the highest-quality models in our experiments. In (c) we illustrate the mask parameter alternation described in \Cref{app:mask-altern} as applied to mask (a), demonstrating that with a full iteration of the algorithm, all links are active at least once. A subset of all available plaquettes are observed to be always frozen or passive (i.e., some surfaces remain colorless), which can be remedied by additionally iterating over the active loop directions. For an interactive version of these figures, see the supplementary jupyter notebook.}
    \label{fig:mask}
\end{figure*}

Once a canonical cell has been chosen, there remains considerable freedom in the parameterization of the map within the cell itself, such as in the choice of coordinate system. While different coordinate choices technically carry the same information, they can result in measurably different model qualities in practice. An intuitive explanation for this observation emerges when discussing the next design choice made in this context, namely, to use two sequential neural splines~\cite{durkan2019neural, Rezende:2020hrd} in order to transform the coordinates. This is precisely the autoregressive transformation of internal degrees of freedom mentioned in \Cref{sec:jacobian}. For simplicity, the splines are always defined on unit intervals with fixed endpoints,
\begin{equation}
    f: [0,1] \longrightarrow [0,1],\quad f(0) = 0,\quad f(1) = 1
\,,
\end{equation}
and we generally employ rational quadratic splines that are strictly monotonic (to ensure invertibility) with parameters determined by context functions. To illustrate the importance of the choice of coordinate system, consider first the cartesian coordinates $(\theta_1,\theta_2)$. One may proceed by computing the re-scaled variable $\theta_1/\hat{\theta}_1 \in [0,1]$ where $\hat{\theta}_1$ is the largest possible value that $\theta_1$ can assume, corresponding to the right-most corner of the canonical cell. Applying the first spline transformation and multiplying the output $f(\theta_1/\hat{\theta}_1) \in [0,1]$ again with $\hat{\theta}_1$, we obtain the transformed variable $\theta_1'$. Fixing $\theta_1'$ in this manner corresponds to drawing a straight vertical line through the canonical cell, thereby constraining the allowed interval for the values that $\theta_2'$ may assume. Hence, applying the second spline now requires both a re-scaling as well as a shift accounting for the offset from the boundaries of the $[0,2\pi] \times [0,2\pi]$ box. As the splines themselves are agnostic to these operations, the resulting ambiguity increases the difficulty of the modeling task. If, on the other hand, one uses polar coordinates $(r,\phi)$ as shown in \Cref{fig:canonical-cell}, no shifts are required and the re-scaling ambiguity is reduced since the allowed maximal values of $r$ (along the bottom right edge of the cell) are tightly constrained. Following this argument, the optimal choice for the origin of the polar coordinate system is the `sharpest' corner of the cell (i.e., the one with the smallest associated angle), since this choice leads to the least ambiguity in the re-scaling of $r$. In general, we find that coordinate systems which minimize these kinds of ambiguities typically result in the highest quality models.

\subsubsection{Masking patterns}\label{sec:spectral-mask}

The optimal choice of masking pattern for a given target density and flow map is not clear a priori. For example, dense masks allow changing all variables in fewer layers than sparse masks, but may remove too much context information. Conversely, sparse masks can enable more expressive conditional transformations simply because more information is available, but may significantly increase the cost of evaluating the model. The testing and comparison of different masks can be automated to some extent by implementing an algorithm that can generate a variety of patterns using only a small number of parameters, ideally applicable to lattices of arbitrary dimensionality. Following the structure of the loop-level coupling layers described above, such an algorithm can be split into two parts. First, a mask distinguishing active from frozen links is generated. Then, from this link mask, a second mask is inferred that distinguishes active, passive, and frozen loops. The basic idea is simple: every loop containing no active link is considered frozen, and the remaining loops are set as either active or passive depending on the number of active links they contain.

\begin{figure}
    \centering
    \subfloat[]{%
        \includegraphics[width=0.67\linewidth]{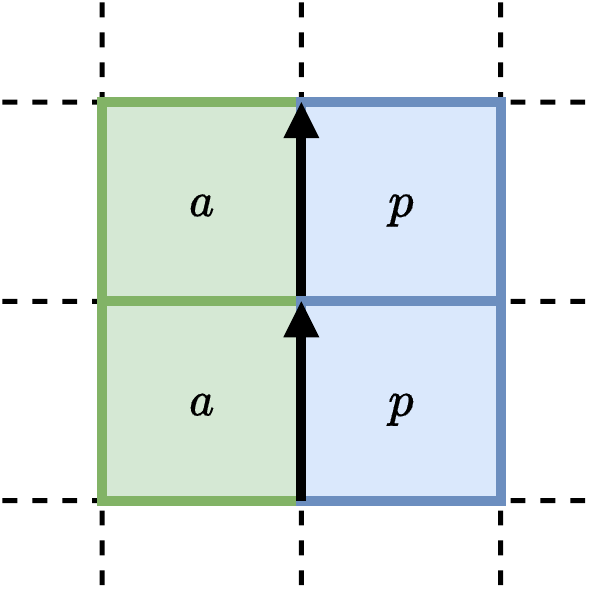}\label{fig:direction-coupling}}\\
    \subfloat[]{%
        \includegraphics[width=0.67\linewidth]{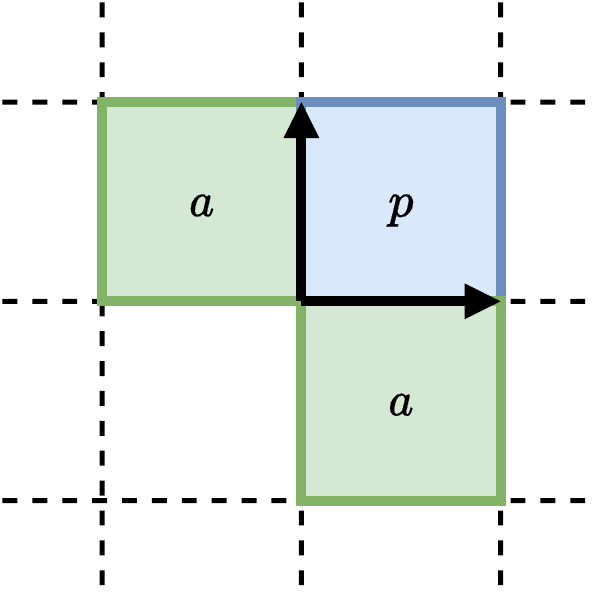}\label{fig:location-coupling}}%
    \caption{Illustration of the difference between the direction (a) and location (b) couplings described in \Cref{sec:spec-dir-loc}, for a two-dimensional lattice. Active and passive plaquettes are marked in green and blue with the labels `a' and `p', respectively, while active links are depicted as black arrows.}
    \label{fig:dir-vs-loc}
\end{figure}

In this work, we demonstrate one instance of an algorithm that meets the above requirements. A number of control parameters are introduced to achieve great flexibility in the types of masks that can be generated. We describe the algorithm in detail in \Cref{app:mask} and provide a concrete implementation in a supplementary jupyter notebook, including interactive versions of the examples of three-dimensional masks shown in \Cref{fig:mask}. The associated parameters are further listed in \Cref{tab:mask-params}. The mask in \Cref{fig:mask2}---which may be considered a generalization of the standard checkerboard---is of particular interest, since every active link is completely surrounded by frozen links. This pattern maximizes the amount of available information in the immediate vicinity of an active link without being excessively sparse. In our experiments, this type of mask typically leads to models of higher quality when compared to, e.g., the `stripe' mask shown in \Cref{fig:mask1}, variants of which were also employed in earlier works~\cite{Kanwar:2020xzo, Boyda:2020hsi, Albergo:2021vyo}.

Masking patterns can be alternated in the sequence of flow layers by looping over permutations of the parameters used to create the initial mask, in a way that generates all unique translations and rotations. Furthermore, it can be advantageous to iterate over all possible choices of active loops for a given set of active links. Two slightly different variants of a suitable mask alternation procedure are described briefly in \Cref{app:mask-altern}, with \Cref{fig:mask-altern} showing that all links are covered in one full iteration of the alternation scheme.

\subsubsection{Direction and location couplings}\label{sec:spec-dir-loc}

Flow layers defined with masks like the ones shown in \Cref{fig:mask} may be called \textit{direction} couplings since all active links are chosen to point in the same direction. Hence, for a particular lattice site, at most one of the associated links is transformed. With sufficient sparsity in the masking pattern, this is a convenient approach to ensuring the consistency of the distinction between active, passive, and frozen loops. However, it is also possible to apply consistent flow transformations simultaneously to all links pointing in all (positive) directions at a given lattice site, since one can always identify a set of independent, non-overlapping active loops for the coupling. Accordingly, these may be called \textit{location} couplings. \Cref{fig:dir-vs-loc} illustrates both possibilities for the two-dimensional case; however, the same construction immediately generalizes to higher dimensions. Appropriate masking patterns for the location coupling can be generated using the same algorithm as for the direction type, with only minor modifications.

Since the two types of couplings differ in their geometry and utilization of frozen information, combining them can be advantageous for the overall expressivity. Indeed, architectures alternating between direction and location couplings lead to improved model qualities in our numerical experiments, in particular when compared to flows consisting of location couplings only.

\subsubsection{Context information}\label{sec:spec-context}

\begin{figure*}
    \includegraphics[width=\textwidth]{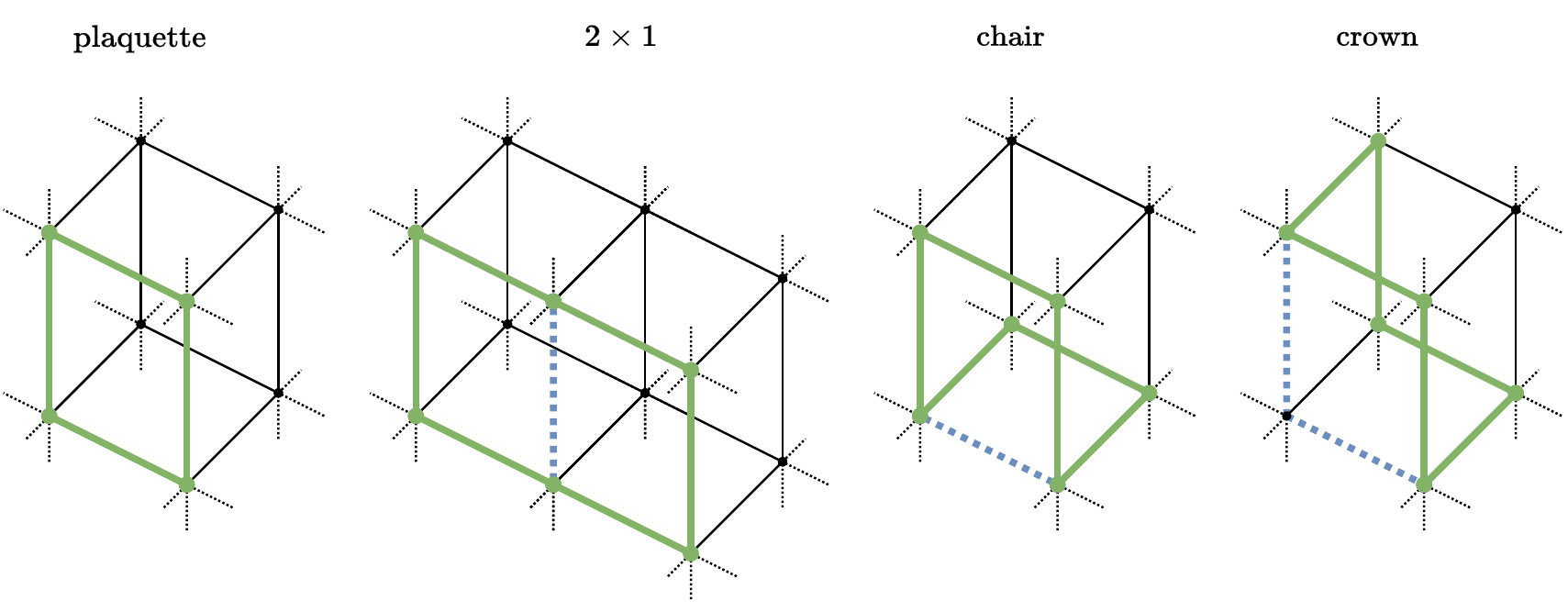}
    \caption{Illustration of possible frozen loops to be used as inputs for the context functions of spectral flows. Dashed blue lines show allowed locations of active links, as long as the associated active loops do not contain any of the frozen links marked in green. For example, the $2\times1$ loop may consist of two passive plaquettes, with the active plaquette attached to the active link but oriented orthogonally to the surface of the loop.}
    \label{fig:frozen-loops}
\end{figure*}

Regarding gauge-invariant context information, in addition to plaquettes we also consider a number of other Wilson loops. This resolves the problem of not being able to construct from only frozen plaquettes an input that contains all available frozen links, since some of these links are only contained in passive plaquettes that also contain active links. Specifically, in addition to plaquettes, we consider $2\times1$ loops, `chairs', and `crowns', as illustrated in \Cref{fig:frozen-loops}. These loop types form a sufficient set for implicitly incorporating the available information from all frozen links. Nevertheless, it may be of additional benefit in practice to also compute more complicated loop shapes explicitly, amounting to pre-processing of the frozen information. In principle, it is also possible to apply arbitrary nested gauge-equivariant transformations~\cite{Favoni:2020reg, Favoni:2022mcg} to the frozen links before using them to compute loops, which may further increase the expressivity of the flow.

Adding the aforementioned loops to the inputs of the context functions generally increases the overall performance in our numerical experiments compared to using plaquettes only, indicating that the expressivity of the flow transformation can indeed be improved by incorporating more complicated loop structures.

\subsection{Link-level flow}\label{sec:link-flow}

An alternative approach to constructing gauge-equivariant coupling layers that enables transformations acting directly on the links can be obtained by generalizing invertible residual networks (i-ResNet)~\cite{pmlr-v97-behrmann19a} to the associated gauge group. Such flows may also be viewed as a generalization of well-known cooling or smoothing algorithms in LGT such as stout smearing, which are directly related to the Wilson or gradient flow~\cite{Bonati:2014tqa}. Hence, they also share some features with continuous formulations of NFs based on ordinary differential equations (ODE flows), which include the original trivializing map construction as a particular instance~\cite{Bacchio:2022vje}. In particular, transformations may be defined via gradients of suitable potential functions, which we discuss in detail in \Cref{sec:residual-flow}. However, in contrast to continuous ODE flows, the architecture is defined in terms of sets of discrete layers whose parameters are not necessarily related by a fictitious flow time, and not all variables are changed in a given layer due to the overall coupling structure. To emphasize the differences in how the compact manifold generalization of coupling layers is achieved in both model families considered in this work, we shall simply refer to them as spectral and residual flows.

The basic i-ResNet architecture utilizes residual blocks of the form
\begin{equation}
    \chi \longmapsto \chi' = f(\chi) = \chi + g(\chi)
\,,
\end{equation}
which can be inverted by fixed-point iteration,
\begin{equation}
    \chi^{(i)} = \chi' - g(\chi^{(i-1)})
\,.
\end{equation}
Hence, the cost of inverting residual blocks is higher than for the forward evaluation, depending on the number of required iterations to achieve a desired precision. Uniqueness of the inverse is guaranteed by satisfying the Lipschitz continuity condition $\mathrm{Lip}(g) < 1$. Similarly, we can define $\rmSU(N)$-residual blocks,
\begin{equation}
    U \longmapsto U' = f(U) \cdot U = e^{\rmi g(U)} U
\,,
\end{equation}
where the multiplication is considered to be element-wise on the link level. They are inverted by the fixed-point iteration
\begin{equation}
    U^{(i)} = e^{-\rmi g(U^{(i-1)})} U'
\,.
\end{equation}
The function $g$ is algebra-valued, $g(U) \in \mathfrak{su}(N)$, and we can write $g = \sum_a g_a T_a$ where the $T_a$ are $N^2 - 1$ Hermitian generators of $\rmSU(N)$ with appropriate normalization---e.g., the Pauli matrices for $N = 2$ or the Gell-Mann matrices for $N = 3$. Implications of the Lipschitz condition on the individual $g_a$ are described in more detail below.

For such residual flows to be gauge-equivariant, under local gauge transformations $f(U)$ must transform like the untraced loops discussed in the previous section. Specifically, under a transformation of $U_\mu(x)$ as defined in \Cref{eq:gauge}, the requirement can be expressed schematically as $[f(U)]_\mu(x) \longmapsto \Omega(x) [f(U)]_\mu(x) \Omega(x)^\dag$. In other words, the value of $f$ for a particular gauge link with specified position $x$ and direction $\mu$ must transform like an untraced loop containing the respective link as its first element; e.g., the plaquette specified by $x,\mu$ with orientation $\nu \neq \mu$. The equivalent constraint on $g$ immediately follows from the usual projection to the algebra and series expansion of the exponential map. It should be emphasized that while similar considerations as for the spectral flow are employed in this construction, the present family of residual flows is defined directly on the gauge links.

\subsubsection{Gradient-based residual flow}\label{sec:residual-flow}

Residual flows with the correct transformation properties may conveniently be defined via gradients of gauge-invariant potentials constructed from traced loops, analogous to a first order finite step size approximation of the well-known Wilson flow. A similar approach based on promoting stout-smearing parameters to learnable weights in order to construct gauge-equivariant neural networks has also been explored~\cite{Nagai:2021bhh, Namekawa:2022liz}, and the present construction may be viewed as a generalization of this ansatz. In the limiting case where the potential is defined solely by traced plaquettes and constant global coefficients---hence, the Wilson action---standard stout smearing is recovered. Importantly, however, by employing conditional transformations based on masking patterns, not only is the Jacobian determinant rendered tractable, but one also opens up the possibility of incorporating context information by dynamically computing coefficients of the potential from frozen loops. It should be noted here that transformations defined via gradients of potentials may already be the most general form of $\rmSU(N)$-residual blocks, similar to the previous statement that all flows for untraced Wilson loops are spectral flows. The semantic distinction is also retained here, in order to account for potentially different implementations that have not yet been considered.

Given a real-valued potential $\phi(U)$, a gradient-based residual block takes the form
\begin{equation}\label{eq:gradient-flow}
    U_\mu(x) \longmapsto e^{\rmi T^a \partial_{x,\mu}^a \phi(U)} U_\mu(x)
\,,
\end{equation}
where the derivative is defined by
\begin{equation}
    \partial_{x,\mu}^a \phi(U) = \left. \frac{d}{ds} \phi(e^{\rmi sT^a_{x,\mu}} U) \right|_{s=0}
\end{equation}
and $T^a_{x,\mu}$ is a site-local generator,
\begin{equation}
    T^a_{x,\mu}(y, \nu) = \begin{cases}
    T^a & (x, \mu) = (y, \nu) \\
    0 & \text{otherwise}
    \end{cases}
\,.
\end{equation}
Any choice of potential defines a map via \Cref{eq:gradient-flow}, but further constraints are required for the map to be a gauge-equivariant diffeomorphism. The equivariance condition implies that the potential itself needs to be gauge-\textit{invariant}. The invertibility requirement corresponds to a Lipschitz condition for the derivatives~\cite{Luscher:2009eq},
\begin{equation}
    \mathrm{Lip}(T^a \partial_{x,\mu}^a \phi) < 1
\,.
\end{equation}
Potentials satisfying the above requirements may be defined in terms of sums of traced loops. With the simplest choice of using only plaquettes, one again recovers the Wilson action,
\begin{equation}
    \phi(U) = c \sum_{x} \sum_{\substack{\mu,\nu \\ \mu < \nu}} \mathrm{Re}\, \mathrm{Tr}\, {P_{\mu\nu}(x)}
\,,
\end{equation}
for some $c \in \mathbb{R}$, thereby making obvious the similarity to stout smearing and the Wilson flow. In Ref.~\cite{Luscher:2009eq}, it was shown that the Lipschitz condition for the above potential assumes the form $\mathrm{Lip}(\phi) < 8c$, implying invertibility of the residual block under the constraint $c < 1/8$.

More generally, essentially arbitrary differentiable functions of both the real and imaginary parts of traces of any set of Wilson loops may be used to define the potential, with appropriate modifications to the Lipschitz condition. A simple choice with analytically computable derivatives is given by polynomials including cross-terms that mix contributions from different loops, promoting the scale factor $c$ to a set of coefficients forming the learnable parameters, which may also be expressed more naturally in terms of a character expansion. Moreover, one may introduce local variations in the parameters by allowing them to depend on gauge-invariant quantities formed by frozen links, which is precisely how context awareness can be implemented here. It is often convenient to compute the Jacobian matrices,
\begin{equation}
    J^{ab}_{\mu\nu}(x, y) T^a_{x,\mu}
    = f(U)^\dag \left. \frac{d}{ds} f(e^{\rmi sT^b_{y,\nu}} U) \right|_{s=0}
\,,
\end{equation}
via automatic differentiation, and evaluate the determinants exactly. Note that this does not reintroduce the unfavorable scaling of the computational cost of the full Jacobian determinant, since the transformations considered here are $\rmSU(N)$-element-wise maps whose internal dimensionality only depends on the properties of the group and is independent of the lattice volume. Automatic differentiation also enables more complicated parameterizations of the potential (such as neural networks) without further conceptual difficulties, simply by applying it twice.

\subsubsection{Masking patterns and context information}\label{sec:residual-mask}

In contrast to spectral flows, the construction of residual flows does not require distinguishing between active, passive, and frozen loops. Instead, one only needs to decompose into active and frozen subsets on the link level, which allows greater flexibility in the types of possible masking patterns. The only requirement is that for the transformation of a particular active link, no other active link is included in the argument of $f$. Hence, the loops used to construct the terms of the gauge-invariant potential must not overlap. This can be easily achieved, e.g., by selecting a particular active link direction (like in the direction coupling described above) combined with a checkerboard masking pattern, and only using terms containing plaquettes. For the frozen information used to compute the coefficients in the potential via context functions, traced loops of in-principle arbitrary shape can be employed as long as they do not contain active links, which can be implemented simply by setting those links to zero in the loop calculation.

\section{Parameterizing and training context functions}\label{sec:context}

In this section, we comment on different possible parameterizations of neural network context functions that exhibit naive scalability and are compatible with both types of flows considered in this work. Furthermore, we discuss optimization strategies and variance reduction techniques for gradient estimates.

\subsection{Neural networks}\label{sec:context-networks}

The choice of neural network parameterization for the context functions affects both the quality of the model as well as the computational cost during training and sampling. Which architecture provides the best trade-off regarding expressivity and efficiency is not clear a priori and must be determined empirically. Following the aforementioned design principle of encoding as many symmetries of the target density as possible into the flow, employing context functions with translation symmetry (including boundary conditions) not only respects this principle, but additionally enables transferring models between different lattice geometries with negligible overhead. We consider three different types of translationally equivariant neural networks, namely standard convolutions, spatially separable convolutions~\cite{1906.06196}, and axial multi-head attention~\cite{1912.12180}. Important aspects of these network types are briefly discussed below.

Pre- and post-processing of context function in- and outputs via standard multilayer perceptrons (MLP) applied locally (i.e., along the channel dimension in the case of convolutions) can further increase the expressivity while preserving translation symmetry. An essential part of any neural network parameterization is also the initial weight distribution, the choice of which can have significant effects on the training dynamics; see also Ref.~\cite{Abbott:2022zsh} for some exploration of this aspect in the present context. Moreover, internally evaluating the networks at lower numerical precision than the flow transformations themselves can decrease the cost while remaining similarly expressive. In general, there may also be constraints on the types of non-linear activation functions one can use in order to respect internal symmetries of the considered field variables. In this work, however, we only consider context functions of gauge-invariant quantities, allowing one to use in-principle arbitrary activations.

\textit{Standard convolutions}: While ubiquitous in modern machine learning applications, default modules for $d > 3$ are currently not available in many popular libraries. In this case, convolutional layers for input data of arbitrary dimension may be constructed recursively from their lower-dimensional counterparts; for pytorch implementations see, e.g., Refs.~\cite{github1, github2, github3}. However, in contrast to the default modules for $d < 4$, their evaluation is less optimized and will likely not exhibit the performance one would expect of a native implementation.

\textit{Spatially separable convolutions}: The idea is to employ a low-rank tensor decomposition of the convolution kernel. For example, a two-dimensional kernel may be represented in terms of an outer product of two vectors. This reduces the number of parameters and makes the evaluation more efficient compared to the standard convolution by requiring fewer arithmetic operations. While this reduces the expressivity in general, in the present context the restriction actually encodes a subgroup---exchange of directions---of the hypercubic symmetry inherent to the lattices considered here, which may be beneficial due to reducing redundancy in the learned representations.

\textit{Axial attention}: Based on self-attention, it is a neural network building block used primarily in transformer-based architectures~\cite{allyouneed}. Self-attention is computationally expensive for multi-dimensional data because the attention is applied to the full flattened input tensor. Axial attention is a more efficient implementation where the attention mechanism is only applied to one axis of a given tensor. In multi-head attention architectures, input data undergoes several independent learned transformations, whose outputs are fed into an attention pooling block. The pooling outputs are then concatenated and used as inputs for another transformation. While the attention mechanism is at the heart of some of the most powerful artificial intelligence applications to date~\cite{Brauwers_2023}, we find that in our experiments, convolution architectures generally train faster and result in higher performances.

\subsection{Optimization procedures}\label{sec:context-optim}

Apart from the design choices for the flow architecture itself, different training strategies and hyperparameter settings can also lead to widely varying final model qualities, convergence properties, and overall training cost. In particular, advanced optimizers implementing adaptive learning rates for individual weights and utilizing momentum information often lead to significant improvements over naive stochastic gradient descent~\cite{1609.04747}. In addition to the choice of the optimizer, decaying the learning rate over the course of training is commonly found to lead to improved model quality and speed of convergence, but requires further tuning. Lowering the learning rate too quickly can prevent the model from maximizing its performance, while a slow decay is computationally inefficient due to plateauing.

In our experiments, Adam~\cite{kingma2017adam} proves to be an effective choice considering both speed of convergence as well as final model quality. When training should be stopped in order to achieve an optimal trade-off between training cost and sampling efficiency depends on the desired level of statistics and the training dynamics. Disentangling all the different contributions to the overall cost of training and sampling is a highly non-trivial task and beyond the scope of this work; see Ref.~\cite{Abbott:2022zsh} for a more detailed analysis.

\subsection{Improving stochastic gradients}\label{sec:context-stoch-grad}

The variance associated with stochastic gradient estimates can be reduced with reasonable overhead by introducing control variates. To this end, first note the following trivial identity for the statistical expectation of a function $f$,
\begin{equation}
\begin{aligned}
    \mathbb{E}[f(x)] &= \mathbb{E}[f(x)] - \alpha \left(\mathbb{E}[g(x)] - C \right) \\
    &= \mathbb{E}[f(x) - \alpha g(x)] + \alpha C
\,,
\end{aligned}
\end{equation}
where we assume that the expectation of the function $g$ is analytically known to be $\mathbb{E}[g(x)] = C$, and $\alpha \in \mathbb{R}$ is a free parameter. It is well known~\cite{variancereduction} that the variance of the resulting unbiased estimator for $\mathbb{E}[f(x)] - \alpha C$ is minimized for the particular choice
\begin{equation}
    \alpha = \frac{\mathrm{Cov}[f(x),g(x)]}{\mathrm{Var}[g(x)]}
\,,
\end{equation}
which can be statistically estimated.

\begin{figure}
    \includegraphics[width=\linewidth]{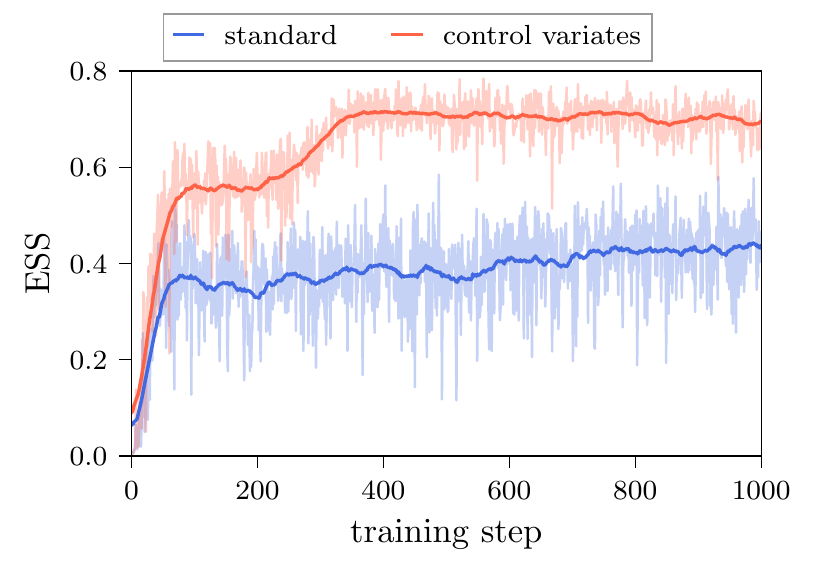}
    \caption{Example comparison of the evolution of the ESS for two equivalent spectral flow architectures with the same random seed trained with and without control variates, demonstrating significantly improved model quality already in the early stages and a pronounced second wind effect. The solid lines show rolling averages with a window size of 50 steps. For reference, the target distribution in this example is LGT with gauge group $\rmSU(3)$ and $\beta = 1$ on a lattice of size $4^4$.}
    \label{fig:control-variates}
\end{figure}

In the present context, we wish to reduce the variance of loss gradient estimates. For the reverse KLD defined in \Cref{eq:kld}, these may be written as
\begin{equation}
    \mathbb{E}[f] \equiv \mathbb{E}_{U \sim q(U)}\left[\frac{\mathrm{d}}{\mathrm{d}\theta} \log \frac{q_\theta(U)}{p(U)}\right]
\,,
\end{equation}
where the subscript indicates that only $q$ depends explicitly on the model weights $\theta$, with all remaining dependency being implicit in the generated samples $U$. Accordingly, we may use the partial derivative as a control variate with vanishing expectation,
\begin{equation}
\begin{aligned}
    C &= \mathbb{E}[g] \equiv \mathbb{E}_{U \sim q(U)}\left[\frac{\partial}{\partial\theta} \log q_\theta(U)\right] \\
    &= \int \mathrm{d}U \, q_\theta \frac{\partial}{\partial\theta} \log q_\theta = \frac{\partial}{\partial\theta} \int \mathrm{d}U \, q_\theta = 0
\,.
\end{aligned}
\end{equation}
In the limit $\alpha \longrightarrow 1$ one recovers the special case of path gradients~\cite{vaitl2022gradients}, and we generally also observe convergence to this limit for long training times. In practice, the procedure requires one additional flow evaluation in the reverse direction per training step. This can be expensive for models with a significant cost asymmetry for forward and reverse evaluation, but the benefits may still outweigh these drawbacks. An example comparison of the evolution of the ESS during training with and without control variates is shown in \Cref{fig:control-variates}.

\section{Numerical demonstration}\label{sec:demo}

For concreteness, we report results from a set of basic numerical experiments for LGT with gauge group $\rmSU(3)$ in four space-time dimensions, demonstrating the implementation of the two families of flows. Both types of transformations can also be freely combined; however, mixed architectures are not considered in this work. Furthermore, a rigorous comparison of the performances at fixed compute budget is not attempted, and differences in the reported ESS values should not be interpreted as benchmarks since they primarily reflect that the two model types are at different stages of optimization and development, and no careful hyperparameter optimization has been performed in either case.

We define a single target by choosing a lattice size of $8^4$ and the Wilson action given in \Cref{eq:action} with $\beta = 1$. An in-depth study of the performance scaling with lattice volume and spacing is not the goal of this work; however, we test the notion that shorter distances in the space of probability densities are easier to model by employing two different prior distributions, one at $\beta = 0$ (corresponding to the Haar measure) and one at $\beta = 0.5$ using the same action as the target, with configurations generated by the heatbath algorithm. At the lattice size considered here, every sample is represented by approximately 300k real numbers. For both models, we follow a mixed-precision approach where the context functions are internally evaluated with single, but the flow transformations and field configurations with double precision.

\subsection{Spectral flow}\label{sec:demo-spectral}

For the spectral flow, we employ a total of 288 layers with alternating direction and location couplings. The eigenvalue spectra are transformed based on the canonicalization approach, using rational quadratic splines with 6 bins to define the mapping of the canonical cell in polar coordinates. For the context functions, we employ standard 4D convolutions constructed by stacking 4 3D convolutions, with a kernel size of 3 in all directions. Each context function features 2 hidden layers with 32 channels and LeakyReLU~\cite{xu2015empirical} activations. The outputs of the convolutional networks are post-processed with MLPs applied along the channel dimension, using 2 hidden layers with 32 neurons and the same activation functions. We use plaquettes as active loops and plaquettes, $2 \times 1$ loops, chairs, and crowns as frozen loops. The parameters for the masking pattern generation are listed in \Cref{tab:mask-params} and are alternated with the second variant of the alternation procedure described in \Cref{app:mask-altern}, with an overall periodicity of 48 layers. We employ the Adam optimizer with a base learning rate of $2\mathrm{e}{-4}$ and hyperparameters $\beta_1 = 0.9$, $\beta_2 = 0.95$, $\epsilon = 1\mathrm{e}{-8}$, using a batch size of 64. The flows are trained for a total of 5k steps, with a reduction of the base learning rate to $1\mathrm{e}{-4}$ after 3k steps and further to $4\mathrm{e}{-5}$ after 4k steps. We use control variates to reduce the variance of loss gradient estimates and additionally clip gradient norms to a maximum value of 50. We evaluate the ESS of the trained models using a batch size of 1024 and obtain $\sim 75\%$ for the $\beta = 0$ and $\sim 82\%$ for the $\beta = 0.5$ prior.

\subsection{Residual flow}\label{sec:demo-residual}

For the residual flow, we employ a total of 128 layers, iterating over the directions $\mu$ of active links and using dense checkerboard masks, thus resulting in an overall periodicity of 8 layers. The gauge-invariant potentials defining the transformation are parameterized with polynomials in the real and imaginary parts of traces of plaquettes containing a given active link, up to second order and including cross terms that mix contributions from different directions. Imposing reflection symmetry for the directions orthogonal to the active link is observed to lead to the best models, compared to both fully isotropic and anisotropic potentials. The derivatives of the potential defining the residual block are computed analytically, while the associated Jacobians are obtained via automatic differentiation. The context functions for the dynamical computation of the coefficients receive all traced rectangular loops up to size $2 \times 2$ as frozen information. The networks are built from 4D separable convolutions, each featuring 2 hidden layers with 32 channels and a kernel size of 3 in all directions, using LeakyReLU activation functions. In each network, no final bias or activation are used and the weights of the last layer are re-scaled by a factor $0.01$ at the beginning of training to initialize the flow close to the identity map. Violations of invertibility are never observed in practice even though the Lipschitz condition is not enforced explicitly. We employ the Adam optimizer with a base learning rate of $1\mathrm{e}{-3}$ and hyperparameters $\beta_1 = 0.9$, $\beta_2 = 0.999$, $\epsilon = 1\mathrm{e}{-8}$, using a batch size of 64. The flows are trained for a total of 2k steps, halving the base learning rate every 500 steps. While control variates can in principle be applied in the same manner as for spectral flows, they are not employed here because the inversion of the residual blocks by fixed-point iteration increases the total cost by roughly an order of magnitude. Whether such an investment is justified by the potential performance gain will be investigated in future work. We evaluate the ESS of the trained models using a batch size of 1024 and obtain $\sim 9\%$ for the $\beta = 0$ and $\sim 18\%$ for the $\beta = 0.5$ prior.

\section{Summary and outlook}\label{sec:summary}

In this work, we report on the development of NF architectures suitable for the application to non-Abelian LGTs in arbitrary space-time dimension. General structural properties and requirements of scalable models are discussed, and implementations in terms of element-wise conditional transformations based on masking patterns (coupling layers) are described in detail. We consider two distinct families of gauge-equivariant maps, namely \emph{spectral} flows autoregressively transforming eigenspectra of untraced Wilson loops, as well as \emph{residual} flows operating directly on gauge links via gradient-based residual blocks. These developments are demonstrated in a proof-of-principle application to $\rmSU(3)$ gauge theory in four space-time dimensions, using a lattice of size $8^4$ and the standard Wilson action with $\beta = 1$ to define the target density. While in-depth benchmarking of these models is beyond the scope of this paper, in addition to the natural extensions of previous work in two dimensions~\cite{Kanwar:2020xzo, Boyda:2020hsi}, the novel developments reported in \Cref{sec:coupling} have proven essential to this demonstration.

In addition to the development of model architectures for arbitrary dimensions, we also report results showcasing the use of flow models within a hybrid approach where conventional MCMC algorithms are used to sample from a non-trivial prior. As anticipated, the model performance improves when the prior density is closer to the target, using $\beta = 0.5$ instead of the Haar measure (corresponding to $\beta = 0$). This is a promising demonstration of the application of NFs to the modeling of maps between densities with similar parameters, which may be particularly valuable if the prior can be chosen such that sequential algorithms like HMC or heatbath decorrelate efficiently, for instance by employing open boundary conditions. On the one hand, this approach sacrifices some attractive advantages of independence sampling, in particular embarrassingly parallel evaluation and direct estimation of thermodynamic observables such as the free energy. However, completely solving the thermodynamic integral defining the partition function is generically more difficult than just estimating expectation values, and is not required for an effective treatment of many ergodicity issues. Hence, a first successful application of flow-based sampling algorithms to LGT at scale is likely going to be based on a hybrid approach.

Our present results constitute an important step in the construction of generative machine learning architectures applicable to physically interesting LGTs, both for generating as well as reweighting statistical ensembles. Apart from potential targets in pure Yang-Mills theory~\cite{Springer:2021liy, Mason:2022trc}, these developments pave the way for an application of NFs to at-scale LQCD calculations in four space-time dimensions, both with direct sampling approaches as well as hybrid algorithms incorporating flow models. To reach this goal, dynamical fermion fields need to be incorporated via flow-based algorithms based on the pseudofermion method, as described in Refs.~\cite{Albergo:2021bna, Abbott:2022zhs}. Once fermionic degrees of freedom have been integrated, engineering and exploration may begin in earnest to develop high-quality flow models for LQCD. First results in this direction, based in part on developments presented here, have recently been reported in Ref.~\cite{Abbott:2022hkm}. If successful, this line of research may enable efficient sampling at small physical lattice spacings where traditional algorithms currently face issues such as critical slowing-down and topological freezing.

\section*{Acknowledgments}

We thank Jakob Finkenrath, Christoph Lehner, and Julian Mayer-Steudte for discussions. RA, GK, DCH, FRL, PES, and JMU are supported in part by the U.S.\ Department of Energy, Office of Science, Office of Nuclear Physics, under grant Contract Number DE-SC0011090. PES is additionally supported by the National Science Foundation under EAGER grant 2035015, by the U.S.\ DOE Early Career Award DE-SC0021006, by a NEC research award, and by the Carl G and Shirley Sontheimer Research Fund. GK is additionally supported by the Schweizerischer Nationalfonds, under grant number 200020\_200424. KC and MSA are supported by the National Science Foundation under the award PHY-2141336. MSA thanks the Flatiron Institute for their hospitality. FRL also acknowledges financial support by the Mauricio and Carlota Botton Fellowship. This work is funded by the U.S.\ National Science Foundation under Cooperative Agreement PHY-2019786 (The NSF AI Institute for Artificial Intelligence and Fundamental Interactions, \url{http://iaifi.org/}). This work is associated with an ALCF Aurora Early Science Program project, and used resources of the Argonne Leadership Computing Facility, which is a DOE Office of Science User Facility supported under Contract DEAC02-06CH11357.

\appendix

\section{Masking pattern algorithm}\label{app:mask}

This appendix provides a detailed description of the algorithm for masking pattern generation discussed in \Cref{sec:spectral-mask}. A python implementation of the full algorithm as well as an interactive version of \Cref{fig:mask} can be found in the supplementary jupyter notebook. The parameter settings used to generate the masks shown in the figure, as well as for the masks employed to obtain the results reported in \Cref{sec:demo-spectral}, are listed in \Cref{tab:mask-params}.

\begin{table*}
    \begin{ruledtabular}
    \begin{tabular}{ccccccc}
    \centering
    mask & dimensions & orientations & width & active phase & orientations to shift & shifts \\ \midrule
    \Cref{fig:mask1} & [6,6,4] & [2] & 4 & 0 & [0,1] & [1,2] \\
    \Cref{fig:mask2} & [6,6,4] & [2] & 4 & 0 & [2,0,1] & [2,1,2] \\
    \Cref{sec:demo-spectral} & [8,8,8,8] & [0] & 4 & 0 & [0,1,2,3] & [2,1,2,2] \\
    \end{tabular}
    \end{ruledtabular}
    \caption{Parameter settings for the masking pattern algorithm used to generate figures and results in this work.}
    \label{tab:mask-params}
\end{table*}

\subsection{Link masks}\label{app:link-mask}

The parameters for the link mask algorithm are:
\begin{itemize}
    \item list \textit{dimensions}: defines the lattice geometry; i.e., how many lattice sites are assumed in each direction. For, e.g., a $8^3 \times 4$ lattice, \textit{dimensions} = [8,8,8,4].
    \item list \textit{orientations}: the directions $\mu$ of active links $U_\mu^A$; e.g.\ for \textit{orientations} = [1,2], active links are subsets of $U_1(x)$ and $U_2(x)$.
    \item int \textit{width}, int \textit{active-phase}: an integer phase is assigned to each link during the construction of the mask; the \textit{width} parameter controls the periodicity of the phases. The value of \textit{active-phase} then determines which links are considered active in a given flow layer to allow for easy mask alternation.
    \item list \textit{orients-to-shift}, list \textit{shifts}: lists of the same length that control the change of phase in each direction when going to the next lattice site.
\end{itemize}
First, a lattice-shaped array \textit{coords} is constructed that contains at each site the coordinate vector of that site. Its shape is equal to \textit{dimensions} with an additional dimension of size $d$. From this array, entries are then selected by \textit{orients-to-shift}, multiplied element-wise with \textit{shifts}, and then summed along the last dimension resulting in an array of shape \textit{dimensions}. The \textit{phases} associated with all lattice sites are obtained as the result of the modulo operation with respect to the \textit{width} parameter. The boolean mask for the sites \textit{site-mask} (an array of shape \textit{dimensions}) is then computed by selecting only the phases equal to \textit{active-phase}. Similarly, the boolean mask for the directions \textit{direction-mask} (a list of length $d$) is constructed by setting entries to True if their associated dimension is contained in \textit{orientations}, and False if not. Finally, the link mask is obtained by element-wise multiplication of the site and direction masks and converting the boolean entries to 0 and 1.

\subsection{Loop masks}\label{app:loop-mask}

The parameters for the loop mask algorithm are:
\begin{itemize}
    \item array \textit{link-mask}: an arbitrary active link mask.
    \item list \textit{loop-types}: the orientations of the active loops; e.g.\ for plaquettes in all directions, in 3D: [01, 02, 12], in 4D: [01, 02, 03, 12, 13, 23].
    \item function \textit{loop-fn}: a function that computes the desired type of loop from a set of links and \textit{loop-types}; e.g.\ all kinds of plaquettes that can be defined for a given lattice geometry.
\end{itemize}
First, from \textit{link-mask} the number of active links \textit{num-links} is counted and their coordinates are stored in \textit{link-coords}. These are then used to split \textit{link-mask} into \textit{num-links} masks of the same shape, one for each active link, which have the value 1 at the position of the associated link and are 0 elsewhere. This split is performed to avoid cancelling out active links of opposite orientation when computing the associated loops. The individual active link masks are stored in the array \textit{split-links} and passed to \textit{loop-fn} together with \textit{loop-types}, resulting in individual loop masks. Taking their absolute value to remove the signs of the active link orientations and adding the individual masks together gives the array \textit{loops} containing one mask of shape \textit{dimensions} for every loop type. It identifies all loops that receive either an active or passive update, i.e.\ all loops containing at least one active link. Accordingly, the boolean array \textit{frozen-loop-mask} identifying all frozen loops can be computed by evaluating where \textit{loops} is zero. 

Then, the array \textit{sign} determining whether a given active link is evaluated forward or backward in the associated loop is computed by passing \textit{link-mask} and \textit{loop-types} to \textit{loop-fn}. By using \textit{link-mask} here, only loops which contain exactly one active link are considered. The \textit{active-loop-mask}, together with \textit{active-loop-ind} identifying the associated loop type, is then computed by iterating \textit{d} over all spatial dimensions and \textit{k} over the number of \textit{loop-types}. In each iteration, one of three possible operations is performed. If \textit{d} is not contained in the \textit{k}th loop type, do nothing. If all elements of the \textit{d}th link mask are zero, append this mask to \textit{active-loop-mask} and -1 to \textit{active-loop-ind}, and skip to the next iteration of \textit{d}. Finally, a \textit{candidate} mask is obtained by element-wise multiplication of the \textit{d}th link mask and the \textit{k}th element of the array \textit{loops} computed previously. If \textit{candidate} is equal to the \textit{d}th link mask, i.e.\ the multiplication left the link mask unchanged and hence the loops of the current type contain active links, multiply \textit{candidate} by the \textit{k}th element of \textit{sign} to set the correct link orientations and remove loops which contain more than one active link. Append the result to \textit{active-loop-mask} and \textit{k} to \textit{active-loop-ind}, and skip to the next iteration of \textit{d}.

\subsection{Mask alternation}\label{app:mask-altern}

An efficient alternation of the algorithm's parameters can be achieved by iterating over different values of \textit{active-phase} with periodicity \textit{width}, thereby generating all unique translations of the given masking pattern. Likewise, all rotations can be generated by additionally iterating over values of the entries of \textit{orientations} and \textit{orients-to-shift} with a periodicity equal to the space-time dimension $d$. Alternatively, a slightly different scheme---employed in Ref.~\cite{Abbott:2022hkm} as well as \Cref{sec:demo-spectral}---can be obtained by setting \mbox{\textit{orientations} = [\textit{orients-to-shift}[0]]} in each iteration.

\bibliographystyle{utphys}
\bibliography{main}

\providecommand{\href}[2]{#2}\begingroup\raggedright\begin{thebibliography}{10}

\bibitem{Luscher:2009eq}
M.~Lüscher \href{http://dx.doi.org/10.1007/s00220-009-0953-7}{{\em Commun.
  Math. Phys.} {\bfseries 293} (2010) 899--919},
  \href{http://arxiv.org/abs/0907.5491}{{\ttfamily arXiv:0907.5491 [hep-lat]}}.

\bibitem{Albergo:2019eim}
M.~S. Albergo, G.~Kanwar, and P.~E. Shanahan
  \href{http://dx.doi.org/10.1103/PhysRevD.100.034515}{{\em Phys. Rev. D}
  {\bfseries 100} no.~3, (2019) 034515},
  \href{http://arxiv.org/abs/1904.12072}{{\ttfamily arXiv:1904.12072
  [hep-lat]}}.

\bibitem{Kanwar:2020xzo}
G.~Kanwar, M.~S. Albergo, D.~Boyda, K.~Cranmer, D.~C. Hackett, S.~Racani\`ere,
  D.~J. Rezende, and P.~E. Shanahan
  \href{http://dx.doi.org/10.1103/PhysRevLett.125.121601}{{\em Phys. Rev.
  Lett.} {\bfseries 125} no.~12, (2020) 121601},
  \href{http://arxiv.org/abs/2003.06413}{{\ttfamily arXiv:2003.06413
  [hep-lat]}}.

\bibitem{Nicoli:2020njz}
K.~A. Nicoli, C.~J. Anders, L.~Funcke, T.~Hartung, K.~Jansen, P.~Kessel,
  S.~Nakajima, and P.~Stornati  (7, 2020) ,
  \href{http://arxiv.org/abs/2007.07115}{{\ttfamily arXiv:2007.07115
  [hep-lat]}}.

\bibitem{Boyda:2020hsi}
D.~Boyda, G.~Kanwar, S.~Racani\`ere, D.~J. Rezende, M.~S. Albergo, K.~Cranmer,
  D.~C. Hackett, and P.~E. Shanahan
  \href{http://dx.doi.org/10.1103/PhysRevD.103.074504}{{\em Phys. Rev. D}
  {\bfseries 103} no.~7, (2021) 074504},
  \href{http://arxiv.org/abs/2008.05456}{{\ttfamily arXiv:2008.05456
  [hep-lat]}}.

\bibitem{Albergo:2021vyo}
M.~S. Albergo, D.~Boyda, D.~C. Hackett, G.~Kanwar, K.~Cranmer, S.~Racani\`ere,
  D.~J. Rezende, and P.~E. Shanahan
  \href{http://arxiv.org/abs/2101.08176}{{\ttfamily arXiv:2101.08176
  [hep-lat]}}.

\bibitem{Albergo:2021bna}
M.~S. Albergo, G.~Kanwar, S.~Racani\`ere, D.~J. Rezende, J.~M. Urban, D.~Boyda,
  K.~Cranmer, D.~C. Hackett, and P.~E. Shanahan
  \href{http://dx.doi.org/10.1103/PhysRevD.104.114507}{{\em Phys. Rev. D}
  {\bfseries 104} no.~11, (2021) 114507},
  \href{http://arxiv.org/abs/2106.05934}{{\ttfamily arXiv:2106.05934
  [hep-lat]}}.

\bibitem{DelDebbio:2021qwf}
L.~Del~Debbio, J.~M. Rossney, and M.~Wilson
  \href{http://arxiv.org/abs/2105.12481}{{\ttfamily arXiv:2105.12481
  [hep-lat]}}.

\bibitem{Hackett:2021idh}
D.~C. Hackett, C.-C. Hsieh, M.~S. Albergo, D.~Boyda, J.-W. Chen, K.-F. Chen,
  K.~Cranmer, G.~Kanwar, and P.~E. Shanahan
  \href{http://arxiv.org/abs/2107.00734}{{\ttfamily arXiv:2107.00734
  [hep-lat]}}.

\bibitem{Nicoli:2021inv}
K.~A. Nicoli, C.~J. Anders, L.~Funcke, T.~Hartung, K.~Jansen, P.~Kessel,
  S.~Nakajima, and P.~Stornati
  \href{http://dx.doi.org/10.22323/1.396.0338}{{\em PoS} {\bfseries
  LATTICE2021} (2022) 338}, \href{http://arxiv.org/abs/2111.11303}{{\ttfamily
  arXiv:2111.11303 [hep-lat]}}.

\bibitem{Foreman:2021ljl}
S.~Foreman, T.~Izubuchi, L.~Jin, X.-Y. Jin, J.~C. Osborn, and A.~Tomiya in {\em
  {38th International Symposium on Lattice Field Theory}}.
\newblock Dec, 2021.
\newblock \href{http://arxiv.org/abs/2112.01586}{{\ttfamily arXiv:2112.01586
  [cs.LG]}}.

\bibitem{Finkenrath:2022ogg}
J.~Finkenrath \href{http://arxiv.org/abs/2201.02216}{{\ttfamily
  arXiv:2201.02216 [hep-lat]}}.

\bibitem{Albergo:2022qfi}
M.~S. Albergo, D.~Boyda, K.~Cranmer, D.~C. Hackett, G.~Kanwar, S.~Racani\`ere,
  D.~J. Rezende, F.~Romero-L\'opez, P.~E. Shanahan, and J.~M. Urban
  \href{http://arxiv.org/abs/2202.11712}{{\ttfamily arXiv:2202.11712
  [hep-lat]}}.

\bibitem{Pawlowski:2022rdn}
J.~M. Pawlowski and J.~M. Urban
  \href{http://arxiv.org/abs/2203.01243}{{\ttfamily arXiv:2203.01243
  [hep-lat]}}.

\bibitem{Gerdes:2022eve}
M.~Gerdes, P.~de~Haan, C.~Rainone, R.~Bondesan, and M.~C.~N. Cheng
  \href{http://arxiv.org/abs/2207.00283}{{\ttfamily arXiv:2207.00283
  [hep-lat]}}.

\bibitem{Singha:2022lpi}
A.~Singha, D.~Chakrabarti, and V.~Arora
  \href{http://arxiv.org/abs/2207.00980}{{\ttfamily arXiv:2207.00980
  [hep-lat]}}.

\bibitem{Abbott:2022zhs}
R.~Abbott {\em et~al.} \href{http://arxiv.org/abs/2207.08945}{{\ttfamily
  arXiv:2207.08945 [hep-lat]}}.

\bibitem{Abbott:2022hkm}
R.~Abbott {\em et~al.} in {\em {39th International Symposium on Lattice Field
  Theory}}.
\newblock 8, 2022.
\newblock \href{http://arxiv.org/abs/2208.03832}{{\ttfamily arXiv:2208.03832
  [hep-lat]}}.

\bibitem{Abbott:2022zsh}
R.~Abbott {\em et~al.} \href{http://arxiv.org/abs/2211.07541}{{\ttfamily
  arXiv:2211.07541 [hep-lat]}}.

\bibitem{Bacchio:2022vje}
S.~Bacchio, P.~Kessel, S.~Schaefer, and L.~Vaitl
  \href{http://arxiv.org/abs/2212.08469}{{\ttfamily arXiv:2212.08469
  [hep-lat]}}.

\bibitem{Komijani:2023fzy}
J.~Komijani and M.~K. Marinkovic
  \href{http://dx.doi.org/10.22323/1.430.0019}{{\em PoS} {\bfseries
  LATTICE2022} (2023) 019}, \href{http://arxiv.org/abs/2301.01504}{{\ttfamily
  arXiv:2301.01504 [hep-lat]}}.

\bibitem{Albandea:2023wgd}
D.~Albandea, L.~Del~Debbio, P.~Hern\'andez, R.~Kenway, J.~M. Rossney, and
  A.~Ramos \href{http://arxiv.org/abs/2302.08408}{{\ttfamily arXiv:2302.08408
  [hep-lat]}}.

\bibitem{Nicoli:2023qsl}
K.~A. Nicoli, C.~J. Anders, T.~Hartung, K.~Jansen, P.~Kessel, and S.~Nakajima
  \href{http://arxiv.org/abs/2302.14082}{{\ttfamily arXiv:2302.14082
  [hep-lat]}}.

\bibitem{R:2023dcr}
D.~P. R \href{http://arxiv.org/abs/2304.01798}{{\ttfamily arXiv:2304.01798
  [hep-lat]}}.

\bibitem{tabak2010}
E.~G. Tabak and E.~Vanden-Eijnden
  \href{http://dx.doi.org/10.4310/CMS.2010.v8.n1.a11}{{\em Commun. Math. Sci.}
  {\bfseries 8} no.~1, (03, 2010) 217--233}.

\bibitem{tabak2013}
E.~G. Tabak and C.~V. Turner
  \href{http://dx.doi.org/https://doi.org/10.1002/cpa.21423}{{\em
  Communications on Pure and Applied Mathematics} {\bfseries 66} no.~2, (2013)
  145--164}.

\bibitem{dinh2015nice}
L.~Dinh, D.~Krueger, and Y.~Bengio
  \href{http://arxiv.org/abs/1410.8516}{{\ttfamily arXiv:1410.8516 [cs.LG]}}.

\bibitem{rezende2016variational}
D.~J. Rezende and S.~Mohamed \href{http://arxiv.org/abs/1505.05770}{{\ttfamily
  arXiv:1505.05770 [stat.ML]}}.

\bibitem{dinh2017density}
L.~Dinh, J.~Sohl-Dickstein, and S.~Bengio
  \href{http://arxiv.org/abs/1605.08803}{{\ttfamily arXiv:1605.08803 [cs.LG]}}.

\bibitem{Lehner:2019wvv}
{USQCD} Collaboration, C.~Lehner {\em et~al.}
  \href{http://dx.doi.org/10.1140/epja/i2019-12891-2}{{\em Eur. Phys. J. A}
  {\bfseries 55} no.~11, (2019) 195},
  \href{http://arxiv.org/abs/1904.09479}{{\ttfamily arXiv:1904.09479
  [hep-lat]}}.

\bibitem{Kronfeld:2019nfb}
{USQCD} Collaboration, A.~S. Kronfeld, D.~G. Richards, W.~Detmold, R.~Gupta,
  H.-W. Lin, K.-F. Liu, A.~S. Meyer, R.~Sufian, and S.~Syritsyn
  \href{http://dx.doi.org/10.1140/epja/i2019-12916-x}{{\em Eur. Phys. J. A}
  {\bfseries 55} no.~11, (2019) 196},
  \href{http://arxiv.org/abs/1904.09931}{{\ttfamily arXiv:1904.09931
  [hep-lat]}}.

\bibitem{Cirigliano:2019jig}
{USQCD} Collaboration, V.~Cirigliano, Z.~Davoudi, T.~Bhattacharya, T.~Izubuchi,
  P.~E. Shanahan, S.~Syritsyn, and M.~L. Wagman
  \href{http://dx.doi.org/10.1140/epja/i2019-12889-8}{{\em Eur. Phys. J. A}
  {\bfseries 55} no.~11, (2019) 197},
  \href{http://arxiv.org/abs/1904.09704}{{\ttfamily arXiv:1904.09704
  [hep-lat]}}.

\bibitem{Detmold:2019ghl}
{USQCD} Collaboration, W.~Detmold, R.~G. Edwards, J.~J. Dudek, M.~Engelhardt,
  H.-W. Lin, S.~Meinel, K.~Orginos, and P.~Shanahan
  \href{http://dx.doi.org/10.1140/epja/i2019-12902-4}{{\em Eur. Phys. J. A}
  {\bfseries 55} no.~11, (2019) 193},
  \href{http://arxiv.org/abs/1904.09512}{{\ttfamily arXiv:1904.09512
  [hep-lat]}}.

\bibitem{Bazavov:2019lgz}
{USQCD} Collaboration, A.~Bazavov, F.~Karsch, S.~Mukherjee, and P.~Petreczky
  \href{http://dx.doi.org/10.1140/epja/i2019-12922-0}{{\em Eur. Phys. J. A}
  {\bfseries 55} no.~11, (2019) 194},
  \href{http://arxiv.org/abs/1904.09951}{{\ttfamily arXiv:1904.09951
  [hep-lat]}}.

\bibitem{Joo:2019byq}
{USQCD} Collaboration, B.~Jo\'o, C.~Jung, N.~H. Christ, W.~Detmold, R.~Edwards,
  M.~Savage, and P.~Shanahan
  \href{http://dx.doi.org/10.1140/epja/i2019-12919-7}{{\em Eur. Phys. J. A}
  {\bfseries 55} no.~11, (2019) 199},
  \href{http://arxiv.org/abs/1904.09725}{{\ttfamily arXiv:1904.09725
  [hep-lat]}}.

\bibitem{Fritzsch:2021klm}
P.~Fritzsch, J.~Bulava, M.~C\`e, A.~Francis, M.~L\"uscher, and A.~Rago
  \href{http://dx.doi.org/10.22323/1.396.0465}{{\em PoS} {\bfseries
  LATTICE2021} (2022) 465}, \href{http://arxiv.org/abs/2111.11544}{{\ttfamily
  arXiv:2111.11544 [hep-lat]}}.

\bibitem{Wilson:1974sk}
K.~G. Wilson \href{http://dx.doi.org/10.1103/PhysRevD.10.2445}{{\em Phys. Rev.
  D} {\bfseries 10} (1974) 2445--2459}.

\bibitem{DUANE1987216}
S.~Duane, A.~Kennedy, B.~J. Pendleton, and D.~Roweth
  \href{http://dx.doi.org/https://doi.org/10.1016/0370-2693(87)91197-X}{{\em
  Physics Letters B} {\bfseries 195} no.~2, (1987) 216--222}.

\bibitem{Creutz:1980zw}
M.~Creutz \href{http://dx.doi.org/10.1103/PhysRevD.21.2308}{{\em Phys. Rev. D}
  {\bfseries 21} (1980) 2308--2315}.

\bibitem{Cabibbo:1982zn}
N.~Cabibbo and E.~Marinari
  \href{http://dx.doi.org/10.1016/0370-2693(82)90696-7}{{\em Phys. Lett. B}
  {\bfseries 119} (1982) 387--390}.

\bibitem{Kennedy:1985nu}
A.~D. Kennedy and B.~J. Pendleton
  \href{http://dx.doi.org/10.1016/0370-2693(85)91632-6}{{\em Phys. Lett. B}
  {\bfseries 156} (1985) 393--399}.

\bibitem{Brown:1987rra}
F.~R. Brown and T.~J. Woch
  \href{http://dx.doi.org/10.1103/PhysRevLett.58.2394}{{\em Phys. Rev. Lett.}
  {\bfseries 58} (1987) 2394}.

\bibitem{Adler:1987ce}
S.~L. Adler \href{http://dx.doi.org/10.1103/PhysRevD.37.458}{{\em Phys. Rev. D}
  {\bfseries 37} (1988) 458}.

\bibitem{Schaefer:2009xx}
S.~Schaefer, R.~Sommer, and F.~Virotta
  \href{http://dx.doi.org/10.22323/1.091.0032}{{\em PoS} {\bfseries LAT2009}
  (2009) 032}, \href{http://arxiv.org/abs/0910.1465}{{\ttfamily arXiv:0910.1465
  [hep-lat]}}.

\bibitem{Metropolis:1953am}
N.~Metropolis, A.~W. Rosenbluth, M.~N. Rosenbluth, A.~H. Teller, and E.~Teller
  \href{http://dx.doi.org/10.1063/1.1699114}{{\em J. Chem. Phys.} {\bfseries
  21} (1953) 1087--1092}.

\bibitem{tierney1994markov}
L.~Tierney {\em the Annals of Statistics} (1994) 1701--1728.

\bibitem{LeCuBottOrrMull9812}
Y.~A. LeCun, L.~Bottou, G.~B. Orr, and K.-R. M{\"u}ller, {\em Efficient
  BackProp}, \href{http://dx.doi.org/10.1007/978-3-642-35289-8_3}{pp.~9--48}.
\newblock Springer Berlin Heidelberg, Berlin, Heidelberg, 2012.
\newblock \url{https://doi.org/10.1007/978-3-642-35289-8_3}.

\bibitem{Kullback:1951}
S.~Kullback and R.~A. Leibler
  \href{http://dx.doi.org/10.1214/aoms/1177729694}{{\em The Annals of
  Mathematical Statistics} {\bfseries 22} no.~1, (1951) 79 -- 86}.

\bibitem{Cotler:2022fze}
J.~Cotler and S.~Rezchikov \href{http://arxiv.org/abs/2202.11737}{{\ttfamily
  arXiv:2202.11737 [hep-th]}}.

\bibitem{Detmold:2021ulb}
W.~Detmold, G.~Kanwar, H.~Lamm, M.~L. Wagman, and N.~C. Warrington
  \href{http://dx.doi.org/10.1103/PhysRevD.103.094517}{{\em Phys. Rev. D}
  {\bfseries 103} no.~9, (2021) 094517},
  \href{http://arxiv.org/abs/2101.12668}{{\ttfamily arXiv:2101.12668
  [hep-lat]}}.

\bibitem{Yunus:2022pto}
C.~Yunus and W.~Detmold
  \href{http://dx.doi.org/10.1103/PhysRevD.106.094506}{{\em Phys. Rev. D}
  {\bfseries 106} no.~9, (2022) 094506},
  \href{http://arxiv.org/abs/2205.01001}{{\ttfamily arXiv:2205.01001
  [hep-lat]}}.

\bibitem{Yunus:2022wdr}
C.~Yunus and W.~Detmold \href{http://dx.doi.org/10.22323/1.396.0145}{{\em PoS}
  {\bfseries LATTICE2021} (2022) 145}.

\bibitem{Rezende:2020hrd}
D.~J. Rezende, G.~Papamakarios, S.~Racani\`ere, M.~S. Albergo, G.~Kanwar, P.~E.
  Shanahan, and K.~Cranmer \href{http://arxiv.org/abs/2002.02428}{{\ttfamily
  arXiv:2002.02428 [stat.ML]}}.

\bibitem{Jansen:2003nt}
K.~Jansen \href{http://dx.doi.org/10.1016/S0920-5632(03)02502-7}{{\em Nucl.
  Phys. B Proc. Suppl.} {\bfseries 129} (2004) 3--16},
  \href{http://arxiv.org/abs/hep-lat/0311039}{{\ttfamily
  arXiv:hep-lat/0311039}}.

\bibitem{Favoni:2020reg}
M.~Favoni, A.~Ipp, D.~I. M\"uller, and D.~Schuh
  \href{http://arxiv.org/abs/2012.12901}{{\ttfamily arXiv:2012.12901
  [hep-lat]}}.

\bibitem{Nagai:2021bhh}
Y.~Nagai and A.~Tomiya \href{http://arxiv.org/abs/2103.11965}{{\ttfamily
  arXiv:2103.11965 [hep-lat]}}.

\bibitem{Namekawa:2022liz}
Y.~Namekawa, K.~Kashiwa, H.~Matsuda, A.~Ohnishi, and H.~Takase
  \href{http://dx.doi.org/10.1103/PhysRevD.107.034509}{{\em Phys. Rev. D}
  {\bfseries 107} no.~3, (2023) 034509},
  \href{http://arxiv.org/abs/2210.05402}{{\ttfamily arXiv:2210.05402
  [hep-lat]}}.

\bibitem{Favoni:2022mcg}
M.~Favoni, A.~Ipp, and D.~I. M\"uller
  \href{http://dx.doi.org/10.1051/epjconf/202227409001}{{\em EPJ Web Conf.}
  {\bfseries 274} (2022) 09001},
  \href{http://arxiv.org/abs/2212.00832}{{\ttfamily arXiv:2212.00832
  [hep-lat]}}.

\bibitem{Aronsson:2023rli}
J.~Aronsson, D.~I. M\"uller, and D.~Schuh
  \href{http://arxiv.org/abs/2303.11448}{{\ttfamily arXiv:2303.11448
  [hep-lat]}}.

\bibitem{Lehner:2023bba}
C.~Lehner and T.~Wettig \href{http://arxiv.org/abs/2302.05419}{{\ttfamily
  arXiv:2302.05419 [hep-lat]}}.

\bibitem{Lehner:2023prf}
C.~Lehner and T.~Wettig \href{http://arxiv.org/abs/2304.10438}{{\ttfamily
  arXiv:2304.10438 [hep-lat]}}.

\bibitem{satorras2022en}
V.~G. Satorras, E.~Hoogeboom, and M.~Welling
  \href{http://arxiv.org/abs/2102.09844}{{\ttfamily arXiv:2102.09844 [cs.LG]}}.

\bibitem{Reh:2023zzh}
M.~Reh, M.~Schmitt, and M.~G\"arttner
  \href{http://arxiv.org/abs/2301.06788}{{\ttfamily arXiv:2301.06788
  [cond-mat.str-el]}}.

\bibitem{wehenkel2020say}
A.~Wehenkel and G.~Louppe \href{http://arxiv.org/abs/2006.00866}{{\ttfamily
  arXiv:2006.00866 [cs.LG]}}.

\bibitem{lee2021universal}
H.~Lee, C.~Pabbaraju, A.~Sevekari, and A.~Risteski
  \href{http://arxiv.org/abs/2107.02951}{{\ttfamily arXiv:2107.02951 [cs.LG]}}.

\bibitem{durkan2019neural}
C.~Durkan, A.~Bekasov, I.~Murray, and G.~Papamakarios in {\em Advances in
  Neural Information Processing Systems}, pp.~7511--7522.
\newblock 2019.
\newblock \href{http://arxiv.org/abs/1906.04032}{{\ttfamily arXiv:1906.04032
  [stat.ML]}}.

\bibitem{pmlr-v97-behrmann19a}
J.~Behrmann, W.~Grathwohl, R.~T.~Q. Chen, D.~Duvenaud, and J.-H. Jacobsen in
  {\em Proceedings of the 36th International Conference on Machine Learning},
  K.~Chaudhuri and R.~Salakhutdinov, eds., vol.~97 of {\em Proceedings of
  Machine Learning Research}, pp.~573--582.
\newblock PMLR, Long Beach, California, USA, 09--15 jun, 2019.
\newblock \url{http://proceedings.mlr.press/v97/behrmann19a.html}.

\bibitem{Bonati:2014tqa}
C.~Bonati and M.~D'Elia
  \href{http://dx.doi.org/10.1103/PhysRevD.89.105005}{{\em Phys. Rev. D}
  {\bfseries 89} no.~10, (2014) 105005},
  \href{http://arxiv.org/abs/1401.2441}{{\ttfamily arXiv:1401.2441 [hep-lat]}}.

\bibitem{1906.06196}
J.~Kossaifi, A.~Toisoul, A.~Bulat, Y.~Panagakis, T.~Hospedales, and M.~Pantic
  \href{http://arxiv.org/abs/1906.06196}{{\ttfamily arXiv:1906.06196 [cs.LG]}}.

\bibitem{1912.12180}
J.~Ho, N.~Kalchbrenner, D.~Weissenborn, and T.~Salimans
  \href{http://arxiv.org/abs/1912.12180}{{\ttfamily arXiv:1912.12180 [cs.CV]}}.

\bibitem{github1}
T.~Gebhard, \url{https://github.com/timothygebhard/pytorch-conv4d}.
\newblock Accessed June 17th, 2022.

\bibitem{github2}
J.~P. Vizcaíno, \url{https://github.com/pvjosue/pytorch_convNd}.
\newblock Accessed June 17th, 2022.

\bibitem{github3}
D.~Boyda, \url{https://github.com/boydad/pytorch_conv4D}.
\newblock Accessed June 17th, 2022.

\bibitem{allyouneed}
A.~Vaswani, N.~Shazeer, N.~Parmar, J.~Uszkoreit, L.~Jones, A.~N. Gomez,
  L.~Kaiser, and I.~Polosukhin {\em CoRR} {\bfseries abs/1706.03762} (2017) ,
  \href{http://arxiv.org/abs/1706.03762}{{\ttfamily 1706.03762}}.
  \url{http://arxiv.org/abs/1706.03762}.

\bibitem{Brauwers_2023}
G.~Brauwers and F.~Frasincar
  \href{http://dx.doi.org/10.1109/tkde.2021.3126456}{{\em {IEEE} Transactions
  on Knowledge and Data Engineering} {\bfseries 35} no.~4, (Apr, 2023)
  3279--3298}, \href{http://arxiv.org/abs/2203.14263}{{\ttfamily
  arXiv:2203.14263 [cs.LG]}}.

\bibitem{1609.04747}
S.~Ruder \href{http://arxiv.org/abs/1609.04747}{{\ttfamily arXiv:1609.04747
  [cs.LG]}}.

\bibitem{kingma2017adam}
D.~P. Kingma and J.~Ba \href{http://arxiv.org/abs/1412.6980}{{\ttfamily
  arXiv:1412.6980 [cs.LG]}}.

\bibitem{variancereduction}
Z.~Botev and A.~Ridder, {\em Variance Reduction}.
\newblock John Wiley \& Sons, Ltd, 2017.

\bibitem{vaitl2022gradients}
L.~Vaitl, K.~A. Nicoli, S.~Nakajima, and P.~Kessel
  \href{http://arxiv.org/abs/2207.08219}{{\ttfamily arXiv:2207.08219 [cs.LG]}}.

\bibitem{xu2015empirical}
B.~Xu, N.~Wang, T.~Chen, and M.~Li, 2015.

\bibitem{Springer:2021liy}
F.~Springer and D.~Schaich \href{http://dx.doi.org/10.22323/1.396.0043}{{\em
  PoS} {\bfseries LATTICE2021} (2022) 043},
  \href{http://arxiv.org/abs/2112.11868}{{\ttfamily arXiv:2112.11868
  [hep-lat]}}.

\bibitem{Mason:2022trc}
D.~Mason, B.~Lucini, M.~Piai, E.~Rinaldi, and D.~Vadacchino
  \href{http://dx.doi.org/10.1051/epjconf/202227408007}{{\em EPJ Web Conf.}
  {\bfseries 274} (2022) 08007},
  \href{http://arxiv.org/abs/2211.10373}{{\ttfamily arXiv:2211.10373
  [hep-lat]}}.

\end{thebibliography}\endgroup

\end{document}